\documentclass{aa}

\usepackage{txfonts}
\usepackage{siunitx}
\usepackage{xurl}
\usepackage{xcolor}
\usepackage{float}

\usepackage{graphicx}   
\graphicspath{ {./figures/} }
\usepackage{bm}
\usepackage{amsmath}
\newcommand{\bnabla}{\bm{\nabla}}
\newcommand{\bcdot}{\bm{\cdot}}

\newcommand{\rmn}[1]{\mathrm{#1}}

\newcommand{\lj}[1]{\textcolor{magenta}}

\usepackage{xpatch}
\usepackage{hyperref}
\hypersetup{
	colorlinks=true,
	breaklinks=true,
	citecolor=blue,
	allcolors=blue,
	frenchlinks=true
}

\makeatletter
\xpatchcmd\NAT@citex
{%
	\@citea\NAT@hyper@{%
		\NAT@nmfmt{\NAT@nm}%
		\hyper@natlinkbreak{\NAT@aysep\NAT@spacechar}{\@citeb\@extra@b@citeb}%
		\NAT@date
	}%
}
{%
	\@citea
	\NAT@nmfmt{\NAT@nm}%
	\NAT@aysep\NAT@spacechar
	\NAT@hyper@{\NAT@date}%
}
{}{}
\xpatchcmd\NAT@citex
{%
	\@citea\NAT@hyper@{%
		\NAT@nmfmt{\NAT@nm}%
		\hyper@natlinkbreak{\NAT@spacechar\NAT@@open\if*#1*\else#1\NAT@spacechar\fi}%
		{\@citeb\@extra@b@citeb}%
		\NAT@date
	}%
}
{
	\@citea
	\NAT@nmfmt{\NAT@nm}%
	\NAT@spacechar\NAT@@open\if*#1*\else#1\NAT@spacechar\fi
	\NAT@hyper@{\NAT@date}%
}
{}{}
\makeatother

\makeatletter
\renewcommand*\aa@pageof{, page \thepage{} of \pageref*{LastPage}}
\makeatother

\graphicspath{/figures/}

\begin{document}

\title{Simulating cosmic ray electron spectra and radio emission from an AGN jet outburst in a cool-core cluster}
 \titlerunning{Simulating cosmic ray electron spectra and radio emission in cluster AGN jets}

\author{L\'{e}na Jlassi\inst{1,2}
	\and
	Rainer Weinberger\inst{1}
	\and
	Christoph Pfrommer\inst{1}
	\and
	Maria Werhahn\inst{3}
	\and
	Joseph Whittingham\inst{1}
	\and
	Philipp Girichidis\inst{4}
}

\institute{Leibniz-Institut f\"{u}r Astrophysik Potsdam (AIP), 
	An der Sternwarte 16, D-14482 Potsdam, Germany\\
	\email{ljlassi@aip.de}
	\and
	Institut für Physik und Astronomie, Universität Potsdam, Karl-Liebknecht-Str. 24/25, 14476 Potsdam, Germany
	\and
	Max-Planck-Institut f\"{u}r Astrophysik , Karl-Schwarzschild-Str. 1, 85748 Garching, Germany
	\and
	Universit\"{a}t Heidelberg, Zentrum f\"{u}r Astronomie, Institut f\"{u}r Theoretische Astrophysik, Albert-Ueberle-Str. 2, 69120 Heidelberg, Germany \label{ITA}
}

\date{\today}

\abstract{
Active galactic nucleus (AGN) powered jets can accelerate cosmic ray electrons, leading to the observed radio synchrotron emission. To simulate this emission, jet dynamics in galaxy clusters must be coupled to electron spectral modelling. We run magneto-hydrodynamic (MHD) simulations of a single AGN jet outburst in a Perseus-like galaxy cluster and adopt a sub-grid model for the acceleration of cosmic ray protons and electrons at unresolved internal shocks in the jet. We evolve cosmic ray electron spectra along Lagrangian trajectories using the Fokker-Planck solver \textsc{Crest} and compute the non-thermal emission using \textsc{Crayon+}. The resulting total electron spectrum reaches a steady-state slope at high momenta, with a gradually decreasing normalization over time, while the lower-momentum portion continues to resemble a freely cooling spectrum. The interaction of the jets with the turbulent cluster environment inflates lobes which rise buoyantly, induce amplification of the magnetic fields and uplift old cosmic ray populations in the wake of the bubbles. We connect radio spectral indices to electron injection ages: at a given radio frequency, weaker magnetic fields are illuminated by higher momenta electrons, whose age is determined by the last injection event. On the other hand, stronger magnetic fields are illuminated by lower momenta electrons, whose age is determined by the maximum energy injection event in the past. This powerful approach allows us to relate the underlying MHD properties to electron spectra and the resulting radio synchrotron emission, thereby enabling us to infer the underlying physics from observed radio properties.}


\maketitle



\section{Introduction}

Radio galaxies and other radio phenomena observed in galaxy clusters (e.g., radio halos, relics) are strong evidence for the existence of relativistic particles experiencing synchrotron acceleration by volume-filling magnetic fields in the intracluster medium (ICM) \citep[see][for reviews on the topic]{vanWeeren2019, Brunetti2014}. Throughout decades of observations, many radio galaxy sub-classes have emerged such as bent-tail radio galaxies \citep{Jones1979, Gendron-Marsolais2020}, X-shape radio galaxies \citep{Yang2019a}, double-double radio galaxies \citep{Schoenmakers2000}, and restarting, dying or remnant radio galaxies \citep{Murgia2011,Brienza2017}. A broader classification from \citet{Fanaroff1974} distinguishes between core-brightened Fanaroff-Riley Type I (FRI), and edge-brightened Fanaroff-Riley Type II (FRII) radio source morphologies \citep{Ledlow1996, Mingo2019}. The subclasses of radio galaxies mentioned above are distinguished based on their environment, physical extent, morphology, distributions of spectral index $\alpha_{\nu}$ (for a synchrotron surface brightness $I_{\nu} \propto \nu^{-\alpha_{\nu}}$, where $\nu$ is the frequency), and radiative ages. The wide range of observed radio galaxy morphologies has expanded our understanding of how AGN jets propagate through and interact with the ICM. However, the physical mechanisms responsible for some of the morphologies observed in these objects remain unclear. Observations of radio galaxies, in combination with theoretical and numerical studies, can also teach us about the physics at play, e.g. how non-thermal particles escaping AGN jets can heat cool-core clusters \citep{Guo2008, Pfrommer2013} or help us pinpoint the dissipation range of hydrodynamical turbulence \citep{Muller2021}. Joint modelling of AGN jet dynamics and their radio emission constitutes a tool that could help tackle the open questions in this field.

Relativistic protons and electrons, so-called cosmic rays (CRs) can be accelerated from a thermal pool of particles by the process of diffusive shock acceleration (DSA) \citep{Axford1977, Bell1978a, Bell1978}, through which particles repeatedly scatter off electromagnetic turbulence across a shock front. These particles can also be re-accelerated by interacting with externally driven turbulence, through a process known as second-order Fermi acceleration \citep{Fermi1949}. They subsequently experience cooling based on the local plasma conditions, and radiate through synchrotron and inverse Compton scattering.

The relativistic nature of AGN jets makes them ideal sites for particle acceleration. It is unclear exactly where and how CRs are accelerated in jets \citep[see][for a recent review]{Matthews2020}, although potential sites include shearing flows \citep{Wang2021, Sironi2021}, reconnection sites \citep{Sironi2014}, or shocks such as re-collimation shocks \citep{Nishikawa2020}, termination shocks \citep{Cerutti2023}, and backflow shocks in the lobe \citep{Matthews2019}. The process responsible for non-thermal particle acceleration hence likely varies depending on the intrinsic jet properties (density, speed, jet magnetization and magnetic field structure), the source power, and the jet launching, collimation and interaction with the ambient medium \citep{Blandford2019}.

While the presence of radio synchrotron points to the presence of CR electrons (CRes) in bubbles inflated by AGN jets, their pressure contribution is not sufficient to explain the total pressure in some observed lobes \citep{Dunn2004, Birzan2008, Croston2018}. For this reason, CR protons (CRps) have been invoked as an additional pressure component. Additionally, CRps can transfer heat to the ICM, providing a means to suppress cooling flows in cool-core clusters \citep[see][for a review]{Ruszkowski2023}. A number of numerical works have thus included CRps in simulations of AGN feedback \citep{Sijacki2008, Guo2011, Ruszkowski2017, Ehlert2018, Wang2020, Su2021, Beckmann2022}.

Simulating the CRe populations responsible for the radio emission in these systems often involves different approaches to those used for CRps, as CRes have cooling times much shorter than CRps \citep{Ensslin2011}. There exists a variety of studies that have tackled the task of modelling CRe injection (or acceleration\footnote{In this paper, we use the terms injection and acceleration interchangeably, whereas other authors \citep[e.g.,][]{Winner2019} distinguish between the two.}) events and their emission, all relying on different models and assumptions. While some relativistic MHD simulations employing Lagrangian tracer particles to evolve the transport equation of CRes are able to resolve various sites of acceleration within the jet structure and simulate non-thermal emission, they are limited to scales of a few kiloparsecs \citep{Vaidya2018, Dubey2023, Mukherjee2021, Meenakshi2024}. On scales of hundreds of kiloparsecs, various works simulate jets and their radio morphology \citep{Horton2020, English2016, Hardcastle2014}, but the method used \citep{Hardcastle2013} does not evolve non-thermal populations including acceleration and radiative processes. One of the earliest works by \citet{Jones2005} developed a method to evolve the distribution function of CRs in time including CR propagation as an active component that interacts with the ICM, later applied to radio galaxies \citep{Mendygral2012, ONeill2019}. Other studies evolve CR spectra, such as \citet{Vazza2021} in a cosmological galaxy cluster, or \citet{Yang2017} to investigate leptonic AGN jets as a mechanism for the formation of the Fermi bubbles \citep{Su2012}. Finally, some works evolve electron Lorentz factors using analytical solutions rather than solving the full transport equation in order to model synchrotron emission \citep{Yates-Jones2022, Jerrim2024, Chen2023}.

The motivation of modelling synthetic radio emission from AGN jets goes beyond studying details related to morphology, interaction with the ambient medium or lobe evolution. AGN jets are thought to play a crucial role in preventing cooling flows in cool-core clusters, a subclass of galaxy clusters with observed cooling times shorter than 1 Gyr \citep{Peterson2006}. The feedback provided by the AGN is perceptible through X-ray cavities excavated by the jets in the ICM \citep{Birzan2004, Rafferty2006}. Radio bubbles are often observed in the same location as these cavities, serving as a direct tracer of the non-thermal jet plasma. Our aim is thus to model the processes leading to the non-thermal emission in the context AGN jet feedback in galaxy clusters, to study the connection between radio and X-ray emission from a theoretical perspective. Our numerical algorithms are consequently developed to be extended to continuous AGN activity in future work.

In this paper, we follow the temporal and spatial evolution of CRe spectra from a single AGN jet outburst in a Perseus-like cool-core cluster, using the Fokker-Planck solver \textsc{Crest} \citep{Winner2019,Winner2020,Whittingham2024}. To do so, we develop sub-grid algorithms for acceleration of CR protons and electrons in our simulations of MHD AGN jets. This paper aims to thoroughly introduce these algorithms and to dissect the spectral evolution of the non-thermal electron population resulting from a single jet outburst of 50 Myr duration. We show what can be learned by connecting CRe spectra and their non-thermal emission using the underlying MHD dynamics.

This paper is organized as follows: in Sect.~\ref{sec:methods}, we describe initial conditions, our jet model and CR implementations. In Sect.~\ref{sec:results}, we describe the results of our algorithms, from electron spectra to radio emission and electron ages. We conclude in Sect.~\ref{sec:conclusions}.

\begin{figure*}[ht]
    \centering
    \includegraphics[width=2\columnwidth]{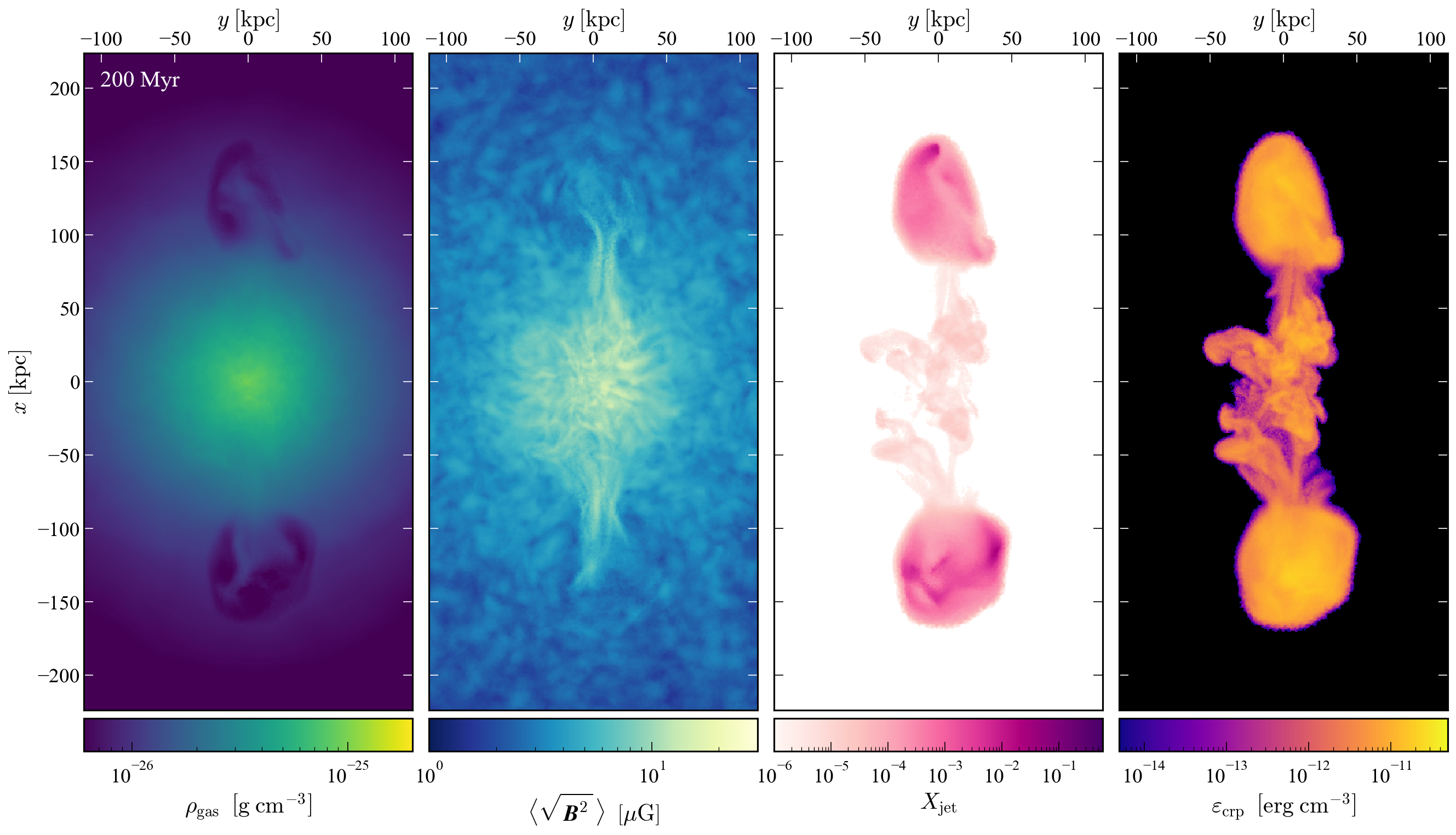}
    \caption{Projections of a single jet outburst in a Perseus-like cluster showing volume-weighted quantities from left to right: mass density, magnetic field, jet mass fraction, CRp energy density. All projections have a depth $\pm$ 60~kpc from the cluster centre. The low-density jet-inflated lobes drag up gas in their wake, amplifying the magnetic field. In contrast to the gas density, in which only the lobes are visible, the mixing of jet material and CRps with the ICM is visible in the central 100 kpc. A movie can be seen \href{https://youtu.be/aMdoZijtoWU}{here}.}
    \label{fig:gas_projections}
\end{figure*}

\section{Methods: simulation setup and algorithms}\label{sec:methods}

We perform 3D MHD simulations using the moving mesh code \textsc{Arepo} \citep{Springel2010, Pakmor2016, Weinberger2020}. The code uses a Voronoi tessellation constructed from mesh-generating points, each point moving close to the local fluid velocity and representing the position of a Voronoi cell, rendering the code quasi-Lagrangian (in turn reducing numerical diffusion). The ideal MHD equations are discretized and solved using a second-order finite-volume Godunov scheme \citep{Pakmor2011, Pakmor2013} which uses an HLLD Riemann solver \citep{Miyoshi2010}. The magnetic divergence constraint $\bm\nabla \bm\cdot \bm{B} = 0$ is maintained using the Powell 8-wave scheme \citep{Powell1999}. This has been demonstrated to be highly robust even in strongly dynamic flows \citep{Pakmor2013, Whittingham2021}.

\subsection{Initial conditions: a Perseus-like galaxy cluster}\label{subsec:ics}

We use initial conditions identical to those described in \citet{Ehlert2023}, of which we summarize the main elements here. We refer the reader to \citet{Ehlert2023} for more details.

We initialize our galaxy cluster in a box of $6000\times6000\times6000 \, \rmn{kpc}^3$ using the radial electron density profile from \citet{Churazov2003} rescaled to a cosmology with $h = 0.67$:
\begin{equation}
\begin{split}
    n_{\rm{e}} & = 46 \times 10^{-3} \left[1 + \left( \frac{r}{60 ~\rmn{kpc}} \right)^2 \right]^{-1.8} \rmn{cm}^{-3} \\
    & + 4.7 \times 10^{-3} \left[ 1 + \left( \frac{r}{210 ~\rmn{kpc}} \right)^2 \right]^{-0.87} \rmn{cm}^{-3},
\end{split}
\end{equation}
set up in hydrostatic equilibrium with a gravitational potential composed of a galaxy cluster potential and a central galaxy potential. The galaxy cluster potential follows an Navarro-Frenk-White (NFW) profile \citep{Navarro1996, Navarro1997} with virial mass $M_{\rmn{200, NFW}} = 8 \times 10^{14} \, \rmn{M}_\sun$ and $R_{\rmn{200,NFW}} = 2$ Mpc and concentration parameter $5$, while the central galaxy potential is modelled as an isothermal sphere with $M_{\rmn{200, ISO}} = 2.4 \times 10^{11}  \, \rmn{M}_\sun$ and $R_{\rmn{200,ISO}} = 15$~kpc \citep{Mathews2006}. The temperature profile is chosen to ensure hydrostatic equilibrium.
Additionally, the initial conditions include turbulent magnetic fields and temperature fluctuations as Gaussian random fields for $\delta T / T$ with dispersion $\sigma = 2$ and mean $\mu = 1$. The resulting power spectrum follows a Kolmogorov slopes on scales smaller than $k_\rmn{inj} = (37.5 \rmn{\,kpc})^{-1}$, and white noise above. We set a constant magnetic-to-thermal pressure ratio of $X_\rmn{B,ICM} = P_{B} / P_\rmn{th} = 0.0125$, where $P_{B}$ and $P_\rmn{th}$ are the magnetic and thermal pressures, respectively. Lastly, we introduce velocity fluctuations as a Gaussian random field within the central $800$~kpc, with a standard deviation $\sigma = 70$~km~s$^{-1}$.

By using turbulent initial conditions, we aim to reproduce the more realistic environment of the cluster core \citep{HitomiCollaboration2018} as well as to break the spherical symmetry of the system. The jet is launched and interacts with the turbulent ICM.

\subsection{AGN jets}\label{subsec:jet}

The AGN jet algorithm is described in detail in \citet{Weinberger2023}. We summarize the main elements of it in this section. We place a black hole particle of mass $M_\rmn{BH} = 4.5 \times 10^9 \, \rmn{M}_{\sun}$ in the centre of the galaxy cluster (i.e., in the centre of the simulation box). We choose a jet luminosity of $L_\rmn{jet} = 3 \times 10^{45}$~erg~s$^{-1}$ for a duration of 50 Myr. The jets are set up within a spherical region with radius $r = 0.6$~kpc. The gas cells within $r$ are initialized with the fiducial low density $\rho_{\rmn{jet}} = 10^{-28} \,$~g~cm$^{-3}$ corresponding to an ICM-to-jet density contrast of $\rho_{\rmn{ICM}} / \rho_{\rmn{jet}} \sim 3 \times 10^3-10^4$. The change in mass and thermal energies required to achieve this low density while maintaining pressure equilibrium between the jet and accretion regions is taken into account to calculate the remaining kinetic energy budget imparted to the jets \citep[see detailed algorithm in][]{Weinberger2023}. The remaining energy is injected as jets into these cells in strictly bipolar directions without an opening angle. Magnetic fields are injected in the jet by setting the magnetic-to-thermal pressure ratio $X_{B, \rm{jet}} = 0.1$ with a toroidal field structure. To trace jet material, we use a scalar $X_{\rm{jet}}$ which is passively advected according to the continuity equation. The cells inside the jet injection region are initialized with $X_{\rm{jet}} = 1$. As the jet mixes with ambient gas cells, $X_{\rm{jet}}$ decreases. The simulation runs for $220$~Myr. This is enough time to follow the evolution of the pair of lobes as they detach and adiabatically expand, and for their corresponding electron population to age.

Cells follow the target gas mass resolution $m_\rmn{target}$ based on their distance from the centre according to
\begin{equation}
    m_\rmn{target} = m_\rmn{target,0} \exp\left({\frac{r}{100 \,\rmn{kpc}}}\right),
\end{equation}
where the target gas mass resolution in our simulations is set to $m_\rmn{target,0} = 1.5 \times 10^6 \, \rmn{M}_\sun$, which is the same resolution level as the {\fontfamily{lmtt}\selectfont Fiducial} simulation in \citet{Ehlert2023}. At a given radial range, we ensure that the gas mass of all Voronoi cells remains within a factor of two of the target mass by explicitly refining and de-refining the mesh cells and also ensure that the volume of adjacent Voronoi cells differs at most by a factor of four. Cells at the outskirts of the galaxy cluster are limited to a maximum volume of $V_\rmn{cell} \sim 370^3 \, $~kpc$^3$. Additionally, cells with $X_{\rm{jet}} > 10^{-3}$ are refined to a target volume $V_\rmn{jet, target} = 0.9$~kpc$^3$, which corresponds to the resolution level of the {\fontfamily{lmtt}\selectfont HR} simulations in \citet{Ehlert2023}. Our simulations therefore include various refinement criteria to adapt the resolution at desired locations.

\subsection{Cosmic ray physics}\label{subsec:cr}

\subsubsection{Cosmic ray protons}\label{subsubsec:crp}

We evolve non-thermal protons in our simulations according to the CR model of \citet{Pfrommer2017} where CRps are treated as a relativistic fluid with adiabatic index $\gamma_{\rmn{cr}} = 4/3$. In our simulations, we do not model CRp losses through hadronic interactions and Alfvén cooling, or transport through streaming and diffusion, making advection the only transport process.

Sites of CR acceleration inside AGN jets include internal shocks at re-collimation sites and backflows, where DSA is able to energize both CRps and CRes as well as Fermi-II acceleration via interactions with MHD turbulence. While the presence of hotspots seen in FRII jets points to termination shocks as the jet encounters the external medium, the absence of these features in FRI jets, nonetheless radio-emitting, points to the existence of shocks inside the jet structure itself. Although the bow shock caused by the propagation of the jet into the ICM is often resolved in our simulations, internal shocks are computationally more challenging to resolve for an AGN jet model aimed at bridging the gap between jet propagation physics and regulating cooling flows \citep[see][for a thorough discussion]{Weinberger2023}. For these reasons, we use a sub-grid prescription for acceleration of CRs.

The total CRp energy is
\begin{equation}\label{eq:fraction_jet_crp}
    E_{\rm{crp}} = \xi_{\rm{crp}} E_{\rm{jet}},
\end{equation}
where $E_{\rm{jet}}$ is the jet energy over a given period of time, and $\xi_{\rm{crp}} = 0.1$ is the efficiency of CRp acceleration. Since the dominant energy injection is in kinetic form, simply setting the jet composition at launching to be a fixed fraction of CR vs. thermal pressure does not result in the desired CRp energy \citep{Weinberger2017}. Replacing a fraction of the kinetic energy with CRp energy, on the other hand, would alter the jet dynamics considerably \citep{Su2021}. To avoid both effects, we do not inject CRs in the jet launching region, but rather convert a fixed fraction of the jet energy from thermal to CRps once the kinetic jet energy has thermalised. Thus, thermal energy is converted into CRp energy inside jet cells over an exponential injection timescale $\tau_{\rmn{inj}} = 10$~Myr. An exponential decay allows for a non-instantaneous acceleration for which a timescale can be chosen, circumventing the issues mentioned above. While the precise value of $\tau_{\rmn{inj}}$ is secondary, this timescale should roughly match the time a Lagrangian element remains in the spine of the jet, because we expect internal acceleration processes to take place over this timescale.

To perform this progressive, exponentially decaying acceleration, the energy $E_{\rm{crp}}$ is distributed amongst cells in the jet region at a given timestep as a passive quantity $\mathcal{E}_{\rmn{crp}}$
\begin{equation}
    E_{\rm{crp}} = \sum_{i\,\in\, \mathrm{jet\, cells}} \mathcal{E}_{\rmn{crp},i} \frac{V_{{\rm{cell}},i}}{V_{\rmn{jet \,region}}},
\end{equation}
where $V_{{\rm{cell}},i}$ is the volume of an individual cell $i$ and $V_{\rmn{jet \,region}}$ is the total volume of the jet region. The associated energy density is evolved according to the continuity equation and allows us to track the energy budget left to be converted in each cell. This allows us to convert a fixed fraction of the total jet energy into CRps, even in cases with varying jet power, e.g., in self-regulated setups \citep{Ehlert2023}.
The change in energy budget $\mathcal{E}_{\rmn{crp}}$ is dictated by an exponential decay,
\begin{equation}
    \frac{\rmn{d}\mathcal{E}_{\rmn{crp}}}{\rmn{d}t} = -\frac{\mathcal{E}_{\rmn{crp}}}{\tau_{\rmn{inj}}}.
\end{equation}
At each timestep, the thermal energy converted into CRps at time $t_{n}$, given by $\Delta E_{\rm{crp}}(t_{i})$, is determined using the current energy budget:
\begin{equation}
    \Delta E_{\rm{crp}}(t_{n}) = \mathcal{E}_{\rm{crp}}(t_{n-1}) \left[1 - \exp{\left(-\frac{\Delta t_{n}}{\tau_{\rmn{inj}}}\right)} \right]
\end{equation}
where $\Delta t_n = t_n - t_{n-1}$. The energy budget tracked by the passive scalar $\mathcal{E}_{\rm{crp}}$ is updated after each timestep by removing the energy just converted into CRp. The energy budget left at time $t_{n}$ is therefore
\begin{equation}
\mathcal{E}_{\rm{crp}}(t_{n}) = \mathcal{E}_{\rm{crp}}(t_0) \exp{\left(\frac{- \sum_{n=1}^{N }  \Delta t_n}{\tau_{\rmn{inj}}}\right)}
\end{equation}
where $N$ is the number of timesteps. In other words, at time $t = \tau_{\rmn{inj}}$ after a specific jet event, we have converted $(1-\rmn{e}^{-1}) = 63\%$ of $E_{\rm{crp}}$, and $(1-\rmn{e}^{-5}) = 99\%$ at $t = 5 \tau_{\rmn{inj}}$. The resulting CRp energy density is shown in the fourth panel of Fig.~\ref{fig:gas_projections}.

We also use this acceleration algorithm for the non-thermal electrons (see Sect.~\ref{subsubsec:cre_acceleration}). Along the Lagrangian trajectory of a CR population, the source function for acceleration is therefore exponentially decreasing with time.

\subsubsection{Cosmic ray electrons}\label{subsubsec:cre}

Even though the \textsc{Arepo} code employs a quasi-Lagrangian moving mesh, the underlying algorithm is a finite-volume approach. Thus, to obtain Lagrangian trajectories, we use tracer particles to follow the evolution of the fluid and its properties in time and space \citep[we use the classical velocity field tracers described in][]{Genel2013}. The CRe population is discretized on these particles, which record fluid properties on the MHD timestep throughout the \textsc{Arepo} simulation. These properties are used by the post-processing \textsc{Crest} code \citep{Winner2019} to evolve the one-dimensional CRe distribution function $f^{\rmn{1D}}(p) = 4 \pi p^2 f^{\rmn{3D}}(p)$, where $f^{\rmn{3D}}(p)$ is the three-dimensional distribution function, according to the Fokker-Planck equation. This is done in normalised momentum space $p = |\bm{p}|  = |\bm{P}| / (m_\rmn{e} c)$ where $m_{\rmn{e}}$ and $c$ are the electron mass and speed of light, respectively. We choose $p$ to range between $10^{-2}$ and $10^{8}$, and use 20 logarithmically-spaced bins per decade.

\textsc{Crest} evolves the distribution function to account for adiabatic changes, Coulomb losses, and radiative losses (inverse Compton, bremsstrahlung, synchrotron). This work introduces a new sub-grid model for injection at AGN jets (Sect.~\ref{subsubsec:cre_acceleration}), which includes a specific treatment of adiabatic changes (Sect.~\ref{subsec:adiabatic_effects}). In principle, \textsc{Crest} is also capable of performing resolved DSA \citep{Whittingham2024} and re-acceleration, Fermi-II momentum diffusion, and sub-grid acceleration at supernovae \citep{Werhahn2025} although we do not use these capabilities in this work. For a detailed description of the physics implemented in the code, see \citet{Winner2019}.

The background CRe energy density field is sampled by the discrete positions of tracer particles. Hence, to recover a continuous energy density field, a Voronoi tessellation is computed in post-processing using the tracer positions as mesh generating points\footnote{The Voronoi mesh constructed from the tracer particle positions in post-processing is different from the MHD mesh constructed on the fly by \textsc{Arepo}.}, whereby each tracer is assigned a volume $V_{\rmn{cell,tr}}$. Initially, all tracer particles have a thermal spectrum (not evolved in \textsc{Crest}). Each tracer particle trajectory leads to an individual tracer spectrum $f(p)$ affected by acceleration and cooling processes. The CRe energy density $\varepsilon_{\rmn{cre}}$ represented by each tracer is computed by integrating the spectrum:
\begin{equation}\label{eq:cre_energy_density}
	\varepsilon_{\rmn{cre}} = \int_0^\infty T_{\rmn{e}}(p) f(p) \rmn{d}p,
\end{equation}
where $T_{\rmn{e}}(p) = (\sqrt{1 + p^2} - 1)m_{\rmn{e}} c^2$ is the CRe kinetic energy. Each tracer particle with a given Voronoi volume $V_{\rmn{cell,tr}}$ therefore represents an energy $E_{\rmn{cre}} = \varepsilon_{\rmn{cre}} V_{\rmn{cell,tr}}$.

To compute cooling processes, we assume a fully ionized primordial gas consisting mostly of hydrogen and helium, with a hydrogen mass fraction $X_\rmn{H} = 0.76$. This corresponds to a mean molecular weight of $\mu = 0.588$ and an electron-to-hydrogen abundance of $x_\rmn{e} = 1.157$. Throughout our work, we assume a redshift of $z = 0$, which corresponds to a cosmic microwave background energy density of $\varepsilon_{\rmn{CMB}} = 4.2 \times 10^{-13}$~erg~cm$^{-3}$ or an equivalent magnetic field strength of $B_{\rmn{CMB}} = 3.2 \, \mu$G.

\subsubsection{Acceleration of CRes in AGN jets}\label{subsubsec:cre_acceleration}

We create tracer particles on-the-fly inside the jet injection region, which subsequently move with the jet fluid. In the simulations presented in this paper, we obtain at least an average of one tracer particle per jet cell with $X_{\rm{jet}} > 10^{-6}$, where $X_{\rm{jet}}$ is the tracer that follows the fraction of jet mass (cf.\ Eq.~\ref{eq:Xjet}).

We make use of the CRp acceleration described in Sect.~\ref{subsubsec:crp} to accelerate CRes. Specifically, the CRe energy density $\varepsilon_{\rm{cre}}$ is a fraction of the energy density injected in CRp $\varepsilon_{\rm{crp}}$:
\begin{equation}\label{eq:fraction_crp_cre}
    \varepsilon_{\rm{cre}} = \xi_{\rm{cre}} \varepsilon_{\rm{crp}},
\end{equation}
where $\xi_{\rm{cre}} = 0.01$ is the efficiency of CRe acceleration.

In order for a tracer particle's spectrum to experience acceleration, we require the CRe population it represents to be inside the jets or the lobes: this corresponds to a jet scalar $X_{\rmn{jet}} > 10^{-3}$, and the gas speed to be above the threshold $\varv_{\rmn{min}} = 3000$~km~s$^{-1}$.

For tracers that fit these criteria, we inject a power law in momentum:
\begin{equation}
	Q_{\rmn{inj}}(p) = r_{\rm{jet}} \frac{C}{\Delta t} p^{-\alpha_{\rmn{inj}}} \Theta (p - p_{\rmn{min}}),
    \label{eq:injected_spectrum}
\end{equation}
where $r_{\rm{jet}}$ is a pre-factor needed to accurately treat adiabatic changes, which is discussed in Sect.~\ref{subsec:adiabatic_effects} and defined in Eq.~\eqref{eq:dilution}, $\Delta t$ is the time difference between two MHD time steps, $\Theta(x)$ is the Heaviside step function, $\alpha_{\rmn{inj}}$ is the spectral index for electron injection, and $p_{\rmn{min}}$ is the minimum injection momentum which fulfils the condition:
\begin{equation}
    \int\limits_{p_\rmn{min}}^{\infty} C p^{-\alpha_\rmn{inj}} T_\rmn{e}(p)\rmn{d}p = \Delta\varepsilon_\rmn{cre},
    \label{eq:p_min}
\end{equation}
where $\Delta\varepsilon_\rmn{cre}$ is the increase in CRe energy density due to acceleration at a given timestep. In practice, $p_{\rmn{min}} \geq 3 \, p_{\rmn{th}}$ where $p_{\rmn{th}}=\sqrt{2 k_B T / (m_\rmn{e} c^2)}$ \citep{Pinzke2013}. The normalization of the spectrum is determined from requiring continuity of the injected CRe population with the Maxwellian 
\begin{equation}\label{eq:normalization_with_thermal_spectrum}
    C = f_{\rmn{th}}(p_{\rmn{min}}) p_{\rmn{min}}^{-\alpha_\rmn{inj}},
\end{equation}
where $f_{\rmn{th}}$ is the thermal Maxwellian. In reality, the spectral index is determined by the compression ratio across the shock, with standard DSA theory predicting a limit of $\alpha_\rmn{inj} = 2$ at strong shocks and $\alpha_\rmn{inj} > 2$ at weaker shocks. The revised theory of non-linear DSA by \citet{Caprioli2019}, which accounts for CR-modified shocks, leads to spectral indices steeper than $\alpha_\rmn{inj} = 2$, even for strong shocks. As discussed previously, we do not resolve internal shocks in the jet and therefore set $\alpha_\rmn{inj} = 2.2$.

A high temperature jet with $T > 10^{11}$~K (see Fig.~2 of \citealt{Ehlert2023}) starts to attain relativistic corrections to MHD. For these reasons, we set the limit $p_{\rmn{min}} \leq 10$ in \textsc{Crest}. Although momenta below $p = 10$ are initially thermal due to the high jet temperatures at injection, this might not hold true at later stages once the jet is no longer active and temperatures decrease \citep{Weinberger2017, Weinberger2023}. Momenta below $p=10$ are populated either through Coulomb cooled populations, adiabatically cooled from higher momenta or through populations injected at lower temperatures, hence with $p_{\rmn{min}} < 10$. For this reason, as well as to ease the comparison of jet, CRp and CRe energies, we integrate over the entire momentum range when calculating CRe energy densities (Eq.~\ref{eq:cre_energy_density}).

\subsection{Adiabatic effects}\label{subsec:adiabatic_effects}

\begin{figure*}[ht]
	\centering
	\includegraphics[width=2.\columnwidth]{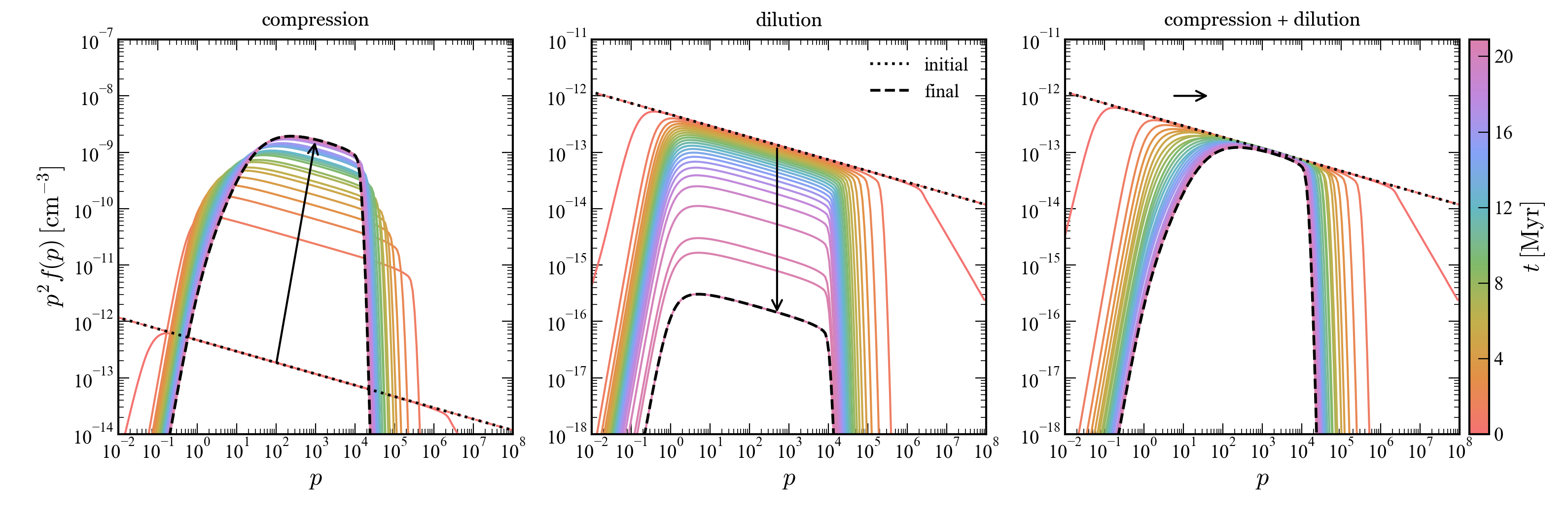}
	\caption{Idealized non-thermal spectra for a single CRe population experiencing compression, dilution, and both simultaneously (in addition to Coulomb and synchrotron/inverse Compton cooling, which narrows the distribution at low and high momenta, respectively). The initial and final CRe distributions are shown as black dotted and dashed lines, respectively. In each case, the vector arrow is calculated based on the expected change in $f(p)$ and $p$ due to changes in the density $\rho_{\rm{gas}}$ and the jet tracer $X_{\rm{jet}}$. 
		\textit{Left:} adiabatic compression of the gas that raises the density by 1000, which leads to an increase in the normalization and a shift in momentum (Eq.~\ref{eq:adiabatic_change}).
		\textit{Middle:} dilution of the CRe population due to a decrease of $X_{\rm{jet}}$ by 1000, which causes a decrease in the normalization of the spectrum (Eq.~\ref{eq:dilution}).
		\textit{Right:} combination of compression and dilution. The density increases by a factor of 1000, and $X_{\rm{jet}}$ is chosen to decrease such that the normalization is left unchanged. In such a scenario, the CRe distribution still moves to the right due to compression.}
	\label{fig:onezone_adiabatic}
\end{figure*}

When unperturbed, the ICM is well described as a stably stratified  atmosphere, with gravity increasing inwards and entropy increasing outwards ($\partial s / \partial r > 0$). The injection of high-temperature, low-density relativistic plasma from AGN jets into the ICM leads to a convectively unstable configuration (Schwarzschild criterion for convective instability)\footnote{Because of magnetic insulation of the bubbles as a result of magnetic draping \citep{Ruszkowski2007,Dursi2008,Pfrommer2010}, heat-flux driven and magneto-thermal instabilities \citep[e.g.,][]{Perrone2022a} are not applicable for the AGN lobes.}, whereby high-entropy fluid is now located in a low-entropy environment. This causes the high-entropy bubbles to rise buoyantly in the cluster atmosphere, which re-establishes a stably stratified atmosphere with entropy increasing with radius. As the AGN bubbles buoyantly rise, they expand and establish pressure equilibrium with the ambient pressure, causing the lobes to adiabatically cool and transferring internal lobe energy to the surrounding ICM. The adiabatic expansion of the AGN lobes and the associated decrease in thermal and CR energy densities is recovered in simulations of AGN jets that include hadronic CR physics \citep{Ruszkowski2017, Ehlert2018, Yang2019a}. However, the impact of adiabatic expansion on the spectrum of leptonic CR populations has yet to be simulated on timescales beyond 100 Myr, when the AGN bubbles have completely detached from the AGN jets and continue rising in the cluster atmosphere.

Using the ideal gas law and assuming a polytropic equation of state, it follows that the final energy density due to change in mass density from $\rho_\rmn{i}$ to $\rho_\rmn{f}$, where the subscripts `i' and `f' correspond to the initial and final values, in the absence of non-adiabatic effects, is $\varepsilon_{\rmn{cr, f}} = \varepsilon_{\rmn{cr,i}} \left( \rho_\rmn{f} / \rho_\rmn{i}\right)^{\gamma_{\rmn{cr}}}$ where $\gamma_{\rmn{cr}} = 4/3$ is the adiabatic index for a relativistic fluid. 

Upon a change in density from $\rho_\rmn{i}$ to $\rho_\rmn{f}$, a typical power-law spectrum of the form given in Eq.~\eqref{eq:injected_spectrum} transforms into
\begin{equation}\label{eq:adiabatic_change}
f_{\rmn{f}}(p) = Cr_{\rho}^{(\alpha_\rmn{inj} +2)/3} p^{-\alpha_\rmn{inj}} \Theta (p - r_{\rho}^{1/3} p_{\rmn{min}}),
\end{equation}
where $r_{\rho} = \rho_\rmn{f} / \rho_\rmn{i}$. This corresponds to a momentum shift from $p_{\rmn{i}}$ to $p_{\rmn{f}} = r_{\rho}^{1/3} p_{\rmn{i}}$, and causes the normalisation to scale according to $r_{\rho}^{(\alpha_\rmn{inj} +2)/3}$ \citep{Ensslin2007}.
The adiabatic expansion of an initial spectrum therefore leads to a final spectrum shifted towards lower momenta and with a lower normalization.

Mixing of low-density jet material with the dense ICM causes the total cell density $\rho_{\rmn{tot}}$ recorded by tracers to increase in time, which adiabatically compresses CRes. On the other hand, this mixing process between the CRe-injected jet material and the ICM gas devoid of CRes, causes CRes to expand into a larger volume. To distinguish adiabatic compression of the CRes from turbulent diffusion due to CRe advection and mixing with the ambient ICM (which keeps the CRe energy invariant), we make use of the jet tracer $X_{\rm{jet}}$, defined as the fraction of the total cell mass $M_{\rm{tot}}$ made up of jet material,
\begin{equation}\label{eq:Xjet}
X_{\rm{jet}} = \frac{m_{\rm{jet}}}{M_{\rm{tot}}},
\end{equation}
where the total cell mass is defined as $M_{\rm{tot}} = m_{\rm{ICM}} + m_{\rm{jet}}$, and $m_\rmn{jet}$ and $m_\rmn{ICM}$ are the masses of the cell made up of jet and non-jet (ICM) material, respectively.

We use $X_{\rm{jet}}$ as a measure of how strongly the initial CRe populations have mixed with the surrounding ICM. Specifically, a decrease in the jet tracer $X_{\rm{jet}}$ corresponds to a smaller mass fraction of jet material, and consequently a `diluted' CRe population, following from the fact that CRe populations are first injected in the jet where $X_{\rm{jet}} = 1$. The CRe number density $n_{\rmn{cre}}$ represented by each tracer is given by
\begin{equation}\label{eq:cre_number_density}
n_{\rmn{cre}} = \int_0^\infty f(p) \rmn{d}p.
\end{equation}
A change in the jet tracer $X_{\rm{jet}}$ encountered along tracer particle trajectories causes the CRe number density to scale according to
\begin{equation}\label{eq:dilution}
n_{\rmn{cre, f}} = r_{\rm{jet}}  n_{\rmn{cre, i}},
\mbox{ where }
    r_{\rm{jet}} = X_{\rm{jet, f}}/X_{\rm{jet, i}}.
\end{equation}

Because the jet tracer $X_{\rmn{jet}}$ is evolved using the continuity equation, it captures the flow of jet material accurately. Combined with the recorded gas density which suffers the consequences of mixing, the jet scalar $X_{\rmn{jet}}$ allows us to account for changes in the number density of CRes. We therefore call this process `dilution'. This corresponds to a change in the normalization of the distribution function $f(p)$ evolved in \textsc{Crest} according to $f_{\rmn{f}} = r_{\rm{jet}} f_{\rmn{i}}$, following Eq.~\eqref{eq:cre_number_density}. Furthermore, due to the continuous nature of our CRe acceleration, we scale the injected source function according to the current dilution state of the gas, which is done through the pre-factor $r_{\rm{jet}}$ seen in Eq.~\eqref{eq:injected_spectrum}.

In Fig.~\ref{fig:onezone_adiabatic}, we show idealized CRe spectra for a single population experiencing compression, dilution, and both simultaneously. In all three cases, we start with a power-law CRe population with slope $\alpha_\rmn{inj} = 2.2$, and choose a magnetic field strength $B = 10 \, \mu \rmn{G}$, a CMB energy density $\varepsilon_{\rmn{CMB}} = 4.2 \times 10^{-13}$~erg~cm$^{-3}$, and gas densities between $\rho_{\rm{gas}} = 10^{-28}\,\rmn{g \, cm}^{-3}$ and $10^{-25}\,\rmn{g \, cm}^{-3}$ (chosen differently for each scenario, as described in the following text). All spectra cool almost immediately  at low momenta up to $p \sim 10^{-1}$ through Coulomb losses, and at high momenta down to $p \sim 10^5$ through synchrotron and inverse Compton emission. CRe cooling timescales for a range of parameters are shown in Fig.~\ref{fig:app_cooling_times} and confirm this, with timescales being as short as 0.1~Myr at the highest and lowest momenta. As time increases, the CRe distribution in all three panels narrows. We only show the spectral evolution for 20~Myrs -- longer timescales would only cause the distribution to narrow further as CRes continue to cool according to the loss timescales shown in Fig.~\ref{fig:app_cooling_times}.
We overplot the initial CRe distribution as a black dotted line and the final CRe distribution as a black dashed line.

In the leftmost panel, CRes experiences adiabatic compression by a factor of $r_{\rho} = 1000$ (from  $\rho =10^{-28}$  to $10^{-25}\,\rmn{g \, cm}^{-3}$), while keeping  $X_{\rm{jet}}$ constant. The arrow shows the expected momentum shift ($\Delta p = 10$) and increase in normalization ($\Delta (p^2f) \approx 1.6 \times 10^4\,$cm$^{-3}$) for such a compression ratio according to Eq.~\eqref{eq:adiabatic_change}. The initial distribution is shifted diagonally upwards and to the right.\footnote{In the case of expansion, the spectrum would shift diagonally downwards and to the left.}

In the middle panel, the CRe population experiences dilution: we model mixing of CRes into a 1000 times larger volume at constant density $\rho_{\rm{gas}} = 10^{-27}\,\rmn{g \, cm}^{-3}$. This decreases $X_{\rm{jet}}$ by a factor of $r_{\rm{jet}} = 1000$ (from $X_{\rm{jet}} = 1$ to $X_{\rm{jet}}=10^{-3}$). The arrows show the expected decrease in normalization for such a dilution event according to Eq.~\eqref{eq:dilution}.
	
Finally, in the rightmost panel, we show what happens when both processes are combined. Specifically, we increase the density by a factor of $r_{\rho} = 1000$ (from  $\rho_{\rm{gas}} = 10^{-28}$  to $10^{-25}\,\rmn{g \, cm}^{-3}$) and decrease the jet tracer such that the change in normalization is balanced by both processes, that is, we require $r_{\rm{jet}} = 1 / r_{\rho}^{1.4} \approx 6 \times 10^{-5}$ for $\alpha_{\rm{inj}} = 2.2$, according to Eqs.~\eqref{eq:adiabatic_change} and \eqref{eq:dilution}. Consequently, there is no change in normalization but only a shift in momentum (as shown by the arrow), which is a consequence of adiabatic compression.

\subsection{Synchrotron emission}\label{subsec:crayon_emission}

We post-process the \textsc{Crest} output using the \textsc{Crayon+} code \citep{Werhahn2021c} which calculates instantaneous non-thermal emission for each CRe population with a given $f(p)$ spectrum. In our case, we are interested in the radio synchrotron emission. The synchrotron emissivity \citep{Rybicki1986} for each tracer particle is given by 
\begin{equation}
    j_\nu = E \frac{\rmn{d}N_\gamma}{\rmn{d}\nu \rmn{d}V \rmn{d}t} = \frac{\sqrt{3} e^3 B_\perp}{m_\rmn{e} c^2} \int^\infty_0 f(p) F(\nu / \nu_c) \rmn{d}p,
\end{equation}
where $N_\gamma$ is the number of photons with energy $E$, $t$ is the unit time, $V$ is the unit volume, $B_\perp$ is the component of the magnetic field perpendicular to the line of sight, and $e$ is the elementary charge. The emissivity $j_\nu$ has units of erg~Hz$^{-1}$~cm$^{-3}$~s$^{-1}$. 
The dimensionless synchrotron kernel $F(x)$ is defined as 
\begin{equation}
    F(x) = x \int^\infty_0 K_{5/3}(\xi) \rmn{d}\xi,
\end{equation}
where $x = \nu / \nu_c$ and $K_{5/3}$ is the modified Bessel function of order 5/3. The calculation of $F(x)$ is done using an analytical approximation \citet{Aharonian2010} for numerical efficiency (for more details, see Appendix~1 in \citealt{Werhahn2021c}).

The critical frequency $\nu_c$ is defined as
\begin{equation}
    \nu_\rmn{c} = \frac{3 e B_\perp \gamma^2}{4 \pi m_\rmn{e} c},
\end{equation}
where $\gamma = \sqrt{1+ p^2}$ is the electron Lorentz factor.
The typical synchrotron emission frequency $v_\rmn{sync} \approx 2 \nu_c$ \citep{Pfrommer2022} is therefore
\begin{equation}\label{eq:nu_sync}
    \nu_{\rmn{sync}} \simeq 1 \: \rm{GHz} \frac{\mathit{B}}{1 \, \mu \rm{G}} \left( \frac{\mathit{\gamma}}{10^4} \right)^2
\end{equation}
where $\gamma \approx p$ for $p \gg 1$. Furthermore, we assume that the synchrotron emission is optically thin to self-absorption, which is appropriate for the ICM. This equation thus allows us to connect the synchrotron emission frequency $\nu_{\rmn{sync}}$ to the momentum of the underlying CRes (we discuss this further in Sect.~\ref{sec:connecting-cre-spectra-and-emission}).

The specific radio synchrotron intensity\footnote{The synchrotron intensity can be converted to units of Jy~arcsec$^{-2}$ by using the conversion 1~Jy~arcsec$^{-2} = 4.25\times10^{-13}$~erg~s$^{-1}$~Hz$^{-1}$~cm$^{-2}$~sterad$^{-1}$.} $I_\nu$ at frequency $\nu$, of units erg~s$^{-1}$~Hz$^{-1}$~cm$^{-2}$~sterad$^{-1}$, is obtained by integrating $j_\nu$ along the line of sight $L$:
\begin{equation}
    I_\nu = \frac{1}{4 \pi} \int^\infty_0 j_\nu \,\rmn{d}L.
\end{equation}
The spectral index is calculated using
\begin{equation}\label{eq:spectral_index}
    \alpha_{\nu_1}^{\nu_2} = - \frac{\log_{10}(I_{\nu_2}/I_{\nu_1})}{\log_{10}(\nu_2 / \nu_1)},
\end{equation}
where $\nu_1 < \nu_2$ such that $\alpha_{\nu_1}^{\nu_2} > 0$ for a synchrotron cooled spectrum where intensity decreases with increasing frequency.

\section{Results}\label{sec:results}

In Fig.~\ref{fig:gas_projections}, we show projections of our MHD simulations at $ t \sim 200$~Myr, that is 150~Myr after the jet has switched off, to emphasize characteristics of the lobe phase. While the lobes are visible as lower-density material in the mass density projection, the morphology of the jet material $X_{\rmn{jet}}$ reveals the larger extent of the lobes. Specifically, the jet material in the wake of the lobes appears perturbed due to mixing of the low-density jet material with the dense ICM core. In the second panel, is it apparent that magnetic fields are amplified in the wake of the rising bubbles (also seen in \citealt{Ehlert2023}), reaching values of 20~$\mu$G. Looking closer at the magnetic field evolution (see \href{https://www.youtube.com/watch?v=43N_huRY__I}{movie}), amplification occurs initially in the jet-ICM interface as the jets expand into the external medium. At a later stage, converging flows compress the gas in the wake of the bubbles. These filaments connect the interior of the bubbles to the ambient ICM so that they also allow CRs to escape and heat the ICM through diffusion and streaming \citep{Ehlert2018}, although the latter are not included in this work for better comparison between jet, CRp and CRe energetics. The fourth panel shows the lobe-filling CRp energy density as a consequence of CR acceleration inside the jet. The mixing of jet material with the ambient medium in the wake of the bubbles (third panel) is reflected in the distribution of CRs, which can be explained by advection of CRs through the turbulent magnetic and velocity fields included in the initial conditions (described in Sect.~\ref{subsec:ics}).

\subsection{Cosmic ray electron spectra}\label{sec:electron-spectra}

\begin{figure*}[ht]
	\centering
	\includegraphics[width=2.\columnwidth]{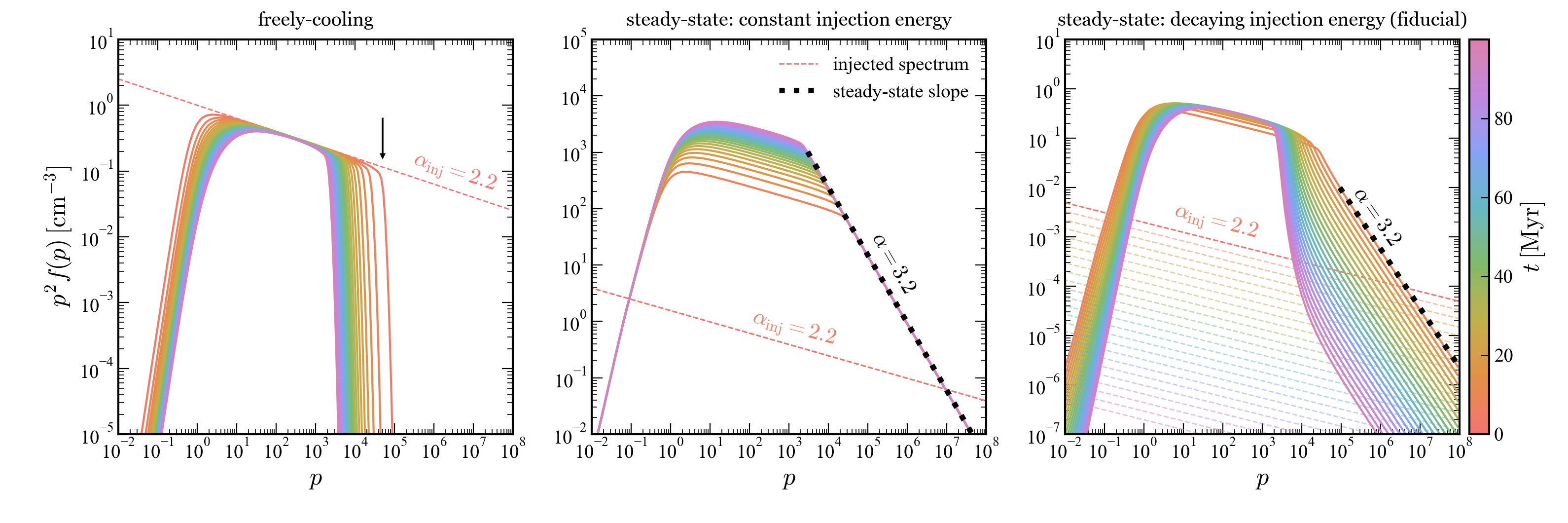}
	\caption{Idealized non-thermal spectra for a single CRe population showing different scenario throughout 100 Myr of evolution. All acceleration events correspond to a power-law CRe population with the same power-law index of $\alpha_{\rmn{inj}} = 2.2$, shown in dashed lines.
		\textit{Left:} freely cooling spectrum--a single acceleration event followed by cooling. The arrow shows the location of the transition momentum between the non-cooled and cooled parts of the spectrum, which shifts to lower momenta over time due to synchrotron losses.
		\textit{Middle:} archetype steady-state spectrum characterized by acceleration with a constant source function and continuous cooling resulting in a steepening of the spectral index by one at high momenta.
		\textit{Right:} acceleration with an exponentially decreasing source function and continuous cooling (fiducial model) results in a freely cooling spectrum except at high momenta where it exhibits a steady state slope (black dotted line) with decreasing normalization.}
	\label{fig:onezone}
\end{figure*}

In order to facilitate the interpretation of CRe spectra in jet-inflated lobes, we first focus on understanding the spectral evolution of a single population of CRes represented by a tracer particle using one zone data (i.e. user-defined, idealised data). We show spectra for three idealized CRe acceleration histories in Fig.~\ref{fig:onezone}. In all three cases, we set the spectral slope of the accelerated spectrum to $\alpha_\rmn{inj} = 2.2$, and use the following values: gas mass density $\rho_{\rm{gas}} = 10^{-27} \, \rmn{g \, cm}^{-3}$, magnetic field strength $B = 10 \, \mu \rmn{G}$ and CMB energy density today of $\varepsilon_{\rmn{CMB}} = 4.2 \times 10^{-13}$~erg~cm$^{-3}$. The dashed lines show the injected power-laws.

The left panel of Fig.~\ref{fig:onezone} shows a single instantaneous acceleration event (shown in the dashed line) followed by continuous cooling. This gives rise to a so-called freely-cooling spectrum. We focus our attention on momenta $p > 10$, specifically on the transition between the flat part of the spectrum (mid-momenta) and the cooled part (high momenta). The momentum at which this transition occurs, shown with a downward arrow, is determined by synchrotron loss timescales. Indeed, for a given magnetic field strength, synchrotron cooling timescales increase with decreasing momentum (cf. Fig.~\ref{fig:app_cooling_times}). This explains why the transition momentum moves towards lower momenta as time increases (red to pink): with increasing time, lower momenta are synchrotron-cooled. At 100~Myr and given a $10 \, \mu \rmn{G}$ magnetic field strength, momenta $p < 10^3$ have thus not yet cooled, as their cooling timescale of $\sim$300~Myr is longer than the current simulation time. For weaker magnetic field strengths, this spectral evolution would be shifted towards higher momenta, with the transition momentum only reaching $p \sim 10^4$ after 100~Myr.

The central panel displays the spectral evolution for a CRe population experiencing continuous and simultaneous acceleration and cooling. We recover a typical steady-state spectrum, whereby the normalization at mid-momenta builds up, while the spectrum at low and high momenta exhibits strong cooling and radiative losses, respectively. The steady-state slope (dotted black line) of $\alpha_{\rmn{steady}} = 3.2$ at high momenta (a steepening by one of the injected power-law index $\alpha_\rmn{inj}=2.2$) is an important characteristic of this spectrum \citep{Sarazin1999}.

Finally, the third panel shows the typical spectral evolution for a single CRe population based on the algorithm described in Sect.~\ref{subsec:cr}: our fiducial model corresponds to continuous injection with an exponentially decreasing source function,\footnote{The chosen injection timescale is $\tau_{\rmn{inj}} = 10$~Myr which is the same throughout this paper.} recorded by tracer particles as they are created and subsequently advected along the jets while they experience various cooling processes. On the one hand, the spectrum displays elements of a freely-cooling spectrum at low- and mid-momenta. On the other hand, as opposed to the middle panel where the source function is constant, the decreasing source function (shown in dashed lines) is unable to build a constant spectrum at high momenta. The steady-state slope of $\alpha_{\rmn{steady}} = 3.2$ is recovered (black dotted line), albeit with a decreasing normalization in the spectrum as the source function decreases. Similarly to the freely-cooling spectrum, the transition momentum between the flat, non-cooled part and the steady-state part of the spectrum moves to the left with increasing time, as lower momenta are synchrotron-cooled.

\begin{figure*}[ht]
	\centering
	\includegraphics[width=1.9\columnwidth]{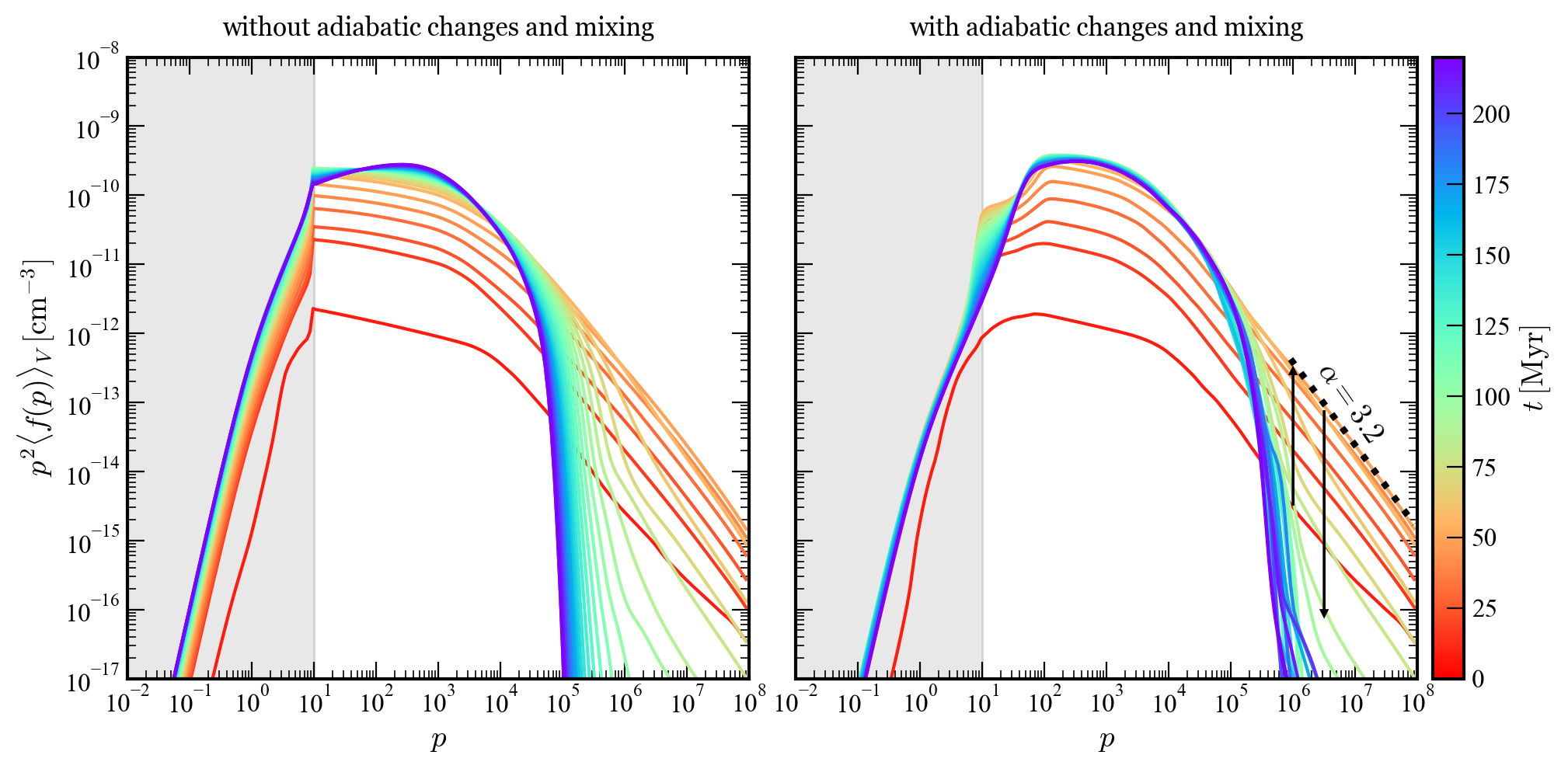}
	\caption{Volume-weighted non-thermal electron spectra throughout 220 Myr of evolution of which the jet is active for the first 50 Myr.
		\textit{Left:} adiabatic changes and mixing effects are switched off and the peak of the spectrum in the mid-momenta range changes relatively little after the jet injection, due to the long cooling times at these energies. 
		\textit{Right:} including adiabatic changes and mixing results in a similar normalization in comparison to the left panel, albeit with a shift of the peak in the first 100~Myr. This can be explained by compression and dilution counteracting each other. The upward and downward arrows show the spectral evolution during and after jet activity, respectively. The spectral index of the accelerated CRes is $\alpha_{\rmn{inj}} = 2.2$. The sharp feature seen at $p = 10$ in the left panel is due to the limit for the minimum injection momentum $p_\rmn{inj,min}$ of the power-law electron population. Although momenta $p < 10$ are hence initially thermal (shown by the greyed out region of the spectrum), they are quickly populated by Coulomb cooling. The steady-state slope (dotted black line) at high momenta is observed in both spectra. }
	\label{fig:cre_spectrum} 
\end{figure*}

In Fig.~\ref{fig:cre_spectrum}, we turn our attention to the temporal evolution of the total electron spectrum from our simulation of a 50~Myr jet outburst (introduced in Fig.~\ref{fig:gas_projections}). We show the spectral evolution with and without adiabatic changes and mixing processes with the ambient ICM. We recover the steady-state slope at high momenta seen in the idealized tests (black dotted line). Within the first 50~Myr, the spectrum increases at high momenta as the populations build up, until a maximum is reached at 50~Myr (shown by the upwards arrow). The spectrum then shows a steady-state behaviour with decreasing amplitude as CRe populations are still accelerated, albeit with lower rates (shown by the downward arrow). Populations at $ 10 < p < 10^4$ are almost unaffected by cooling, which is expected for magnetic field strengths $B \sim 1\,\mu \rm{G}$ and mass densities $\rho_{\rm{gas}} \sim 10^{-27}\,\rmn{g \, cm}^{-3}$, which correspond to cooling times longer than 300~Myr, as shown in Fig.~\ref{fig:app_cooling_times}. This total spectrum, however, combines all individual tracer spectra, each shaped by different magnetic field strengths between $0.01 \leq B/\mu\rmn{G} \leq 30$ and densities between $10^{-29} \leq \rho_{\rm{gas}}/ \rm{g\,cm}^{-3}\leq 10^{-25}$ encountered along their trajectory. This distribution of Lagrangian histories causes a curved shape of the total spectrum at $10<p<10^4$, as opposed to the flatter spectral shape observed at those momenta in the single particle spectra of Fig.~\ref{fig:onezone}.

Comparing the earliest spectra (red curve with the lowest normalization) with and without adiabatic changes and mixing, the peak of the distribution at $p \sim 10$ in the left panel has already shifted rightwards to lie around $p \sim 10^2$ in the model with adiabatic changes and mixing, which indicates an overall adiabatic compression of the CRe population. However, because the normalization of the distribution is nearly unchanged ($\sim 2 \times 10^{-12}\,$cm$^{-3}$), this can only be explained by an additional dilution of the CRe population, as we showed in Fig.~\ref{fig:onezone_adiabatic} with idealized spectra. The increase in normalization caused by compression has been balanced by the decrease in normalization due to dilution of CRes created in the jet and mixing with the ambient ICM. The peak of this earliest spectrum also highlights the minimum injection momentum $p_\rmn{min}$ of the injected power-law electron distributions, which is high due to the high temperatures in the jet, following Eq.~\eqref{eq:normalization_with_thermal_spectrum}. The upper limit we impose on the minimum injection momentum at $p_\rmn{min} = 10$ explains the sharp feature seen in the spectrum without adiabatic changes and mixing, which is smoothed on the right-hand panel due to adiabatic and mixing effects.
	
For both models, the normalization of the peak of the distribution is nearly unchanged in the last 100~Myr, and is almost identical between both models ($p^2\langle f(p) \rangle_{V} \approx 2-3 \times 10^{-10}\,$cm$^{-3}$). Our low-density, FRI-like jets reveal that CRes experience compression simultaneously with dilution, as they mix with the external medium, causing the gas density to increase, and the jet tracer to decrease. This leaves the total electron spectrum at $p > 10^2$ almost identical to a case where adiabatic and mixing effects are absent. Nonetheless, individual CRe populations probed in physical space (Sect.~\ref{sec:connecting-cre-spectra-and-emission}) show variations depending on their individual histories.

Finally, we turn our attention to the feature seen in the model with adiabatic changes and mixing at around $6 \leq p \leq 60$, which is absent in the left panel. Over time, the bump peaking at $p \sim 10$ (red line) progressively dampens to produce an increasing spectrum at low momenta (purple line), reminiscent of a Coulomb cooled spectrum. However, this feature is a pure consequence of adiabatic changes, as it is also present in our model where CRes are modelled with only adiabatic terms (not shown).	This feature, occurring in the last $\sim$100~Myr, is reflected at high momenta ($p \sim 10^6$) where the spectral shape displays more horizontal scatter, causing the spectral lines at different times to overlap. On the other hand, the model without adiabatic changes and mixing shows a spectral evolution typical of cooling CRes, exhibited by a decreasing transition momentum from $p=10^6$ to $p=10^5$ between the non-cooled and cooled parts of the spectrum (blue to purple curves). Both of these features hint at compression taking place, which moves the spectrum towards the right. This effect is also visible in projections of the CRe energy density $\varepsilon_{\rm{cre}}$ in Appendix~\ref{sec:appendix_CRep_distributions} comparing various \textsc{Crest} models. Specifically, Fig.~\ref{fig:app_crep_projections} indicates that CRes in the wake of the bubbles are compressed.

\begin{figure}[ht]
	\centering
	\includegraphics[width=1\columnwidth]{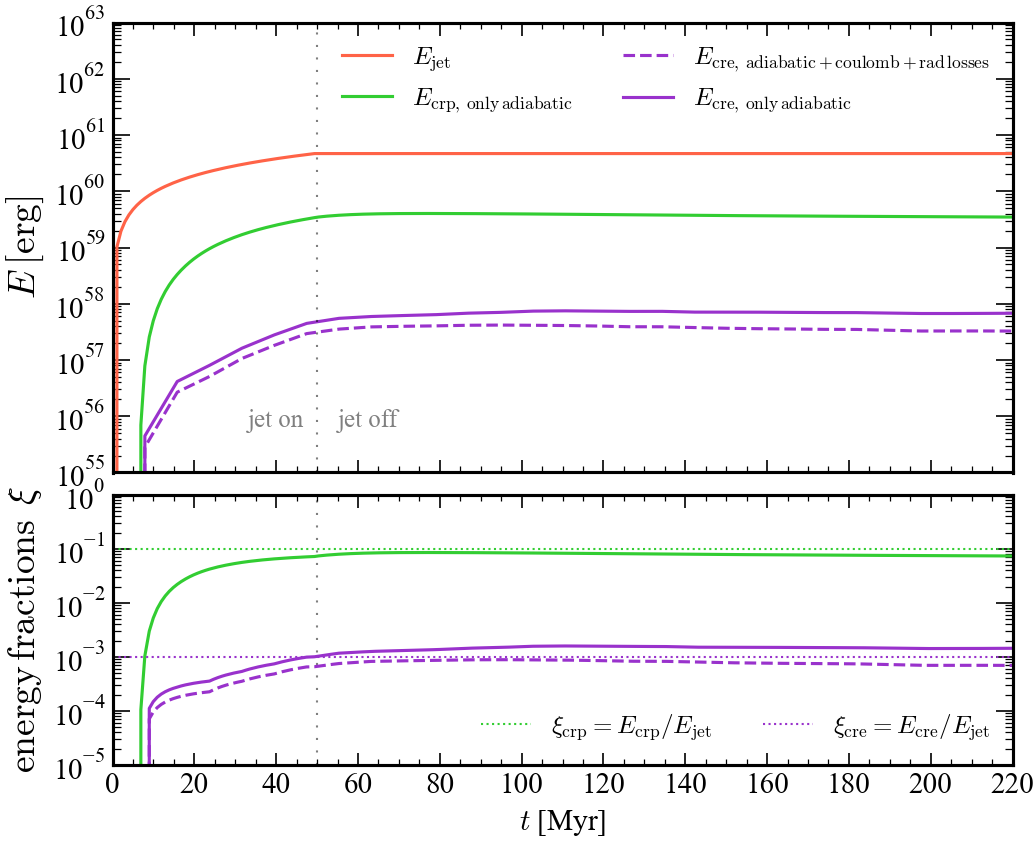}
	\caption{\textit{Top row:} Time evolution of the jet, CRp and CRe energies. The latter is shown with two models: one with only adiabatic and mixing terms, and one with adiabatic and mixing effects, Coulomb and radiative losses.
		\textit{Bottom row:} Time evolution of the energy fractions $\xi$ between jet, CRp and CRe energies. We recover the energy fractions in our simulation output that were chosen as parameters (see Eqs.~\ref{eq:fraction_jet_crp} and~\ref{eq:fraction_crp_cre}), confirming that our acceleration algorithms behave as expected, with minor deviations for the CRe energy due to discretization effects (see text and Appendix~\ref{sec:appendix_CRep_distributions} for details).}
	\label{fig:energies}
\end{figure}

As described in Sect.~\ref{subsubsec:cre}, we can calculate the CRe energy $E_{\rmn{cre}}$ by integrating CRe spectra and summing over the entire volume. In Fig.~\ref{fig:energies}, we show the evolution of the cumulative jet, CRp and CRe energies. This allows us to confirm the energetics of our CR acceleration models. For this specific purpose, we show the evolution of the jet-to-CRp and jet-to-CRe energy fractions, and compare them to the values chosen as model parameters (see Eqs.~\ref{eq:fraction_jet_crp} and \ref{eq:fraction_crp_cre}) in the bottom panel. During jet activity, all energies increase, reaching a plateau shortly after the jet switches off. The CRp and CRe energies stabilize slightly after the jet switches off, due to the progressive, exponential nature of our acceleration algorithm. We successfully recover total CRp and CRe energies in agreement with the chosen conversion efficiencies between jet and CRps, $\xi_{\rm{crp}} = 0.1$, and jet and CRes $\xi_{\rm{cre}} = 0.01$, as shown in the lower panel. As discussed in Sect.~\ref{subsubsec:crp}, converting a fixed fraction of jet energy into CRs without altering jet dynamics is not trivial. Additionally, due to adiabatic processes, we do not expect the simulation to exactly match the input parameters. Indeed, the CRp energy fraction is marginally lower than the parameter value, indicating an overall expansion. Interestingly, however, the CRe energy evolved with only adiabatic and mixing processes is marginally larger than the parameter value. This can be explained by a difference in numerical discretization between the velocity field tracers and the background mesh in converging or diverging flows with large values of $|\bnabla\bcdot \bm{\varv}|$ \citep{Genel2013}, which we discuss in greater detail in Appendix~\ref{sec:appendix_CRep_distributions}. Including Coulomb and radiative losses (dashed purple line) has an immediate effect on the CRe energy even during jet activity, due to the fast cooling losses (less than 10~Myrs at low and high momenta, cf.\ Fig.~\ref{fig:app_cooling_times}). Overall, these loss terms reduce the total CRe by a factor of two.

\begin{figure}[ht]
	\centering
	\includegraphics[width=1\columnwidth]{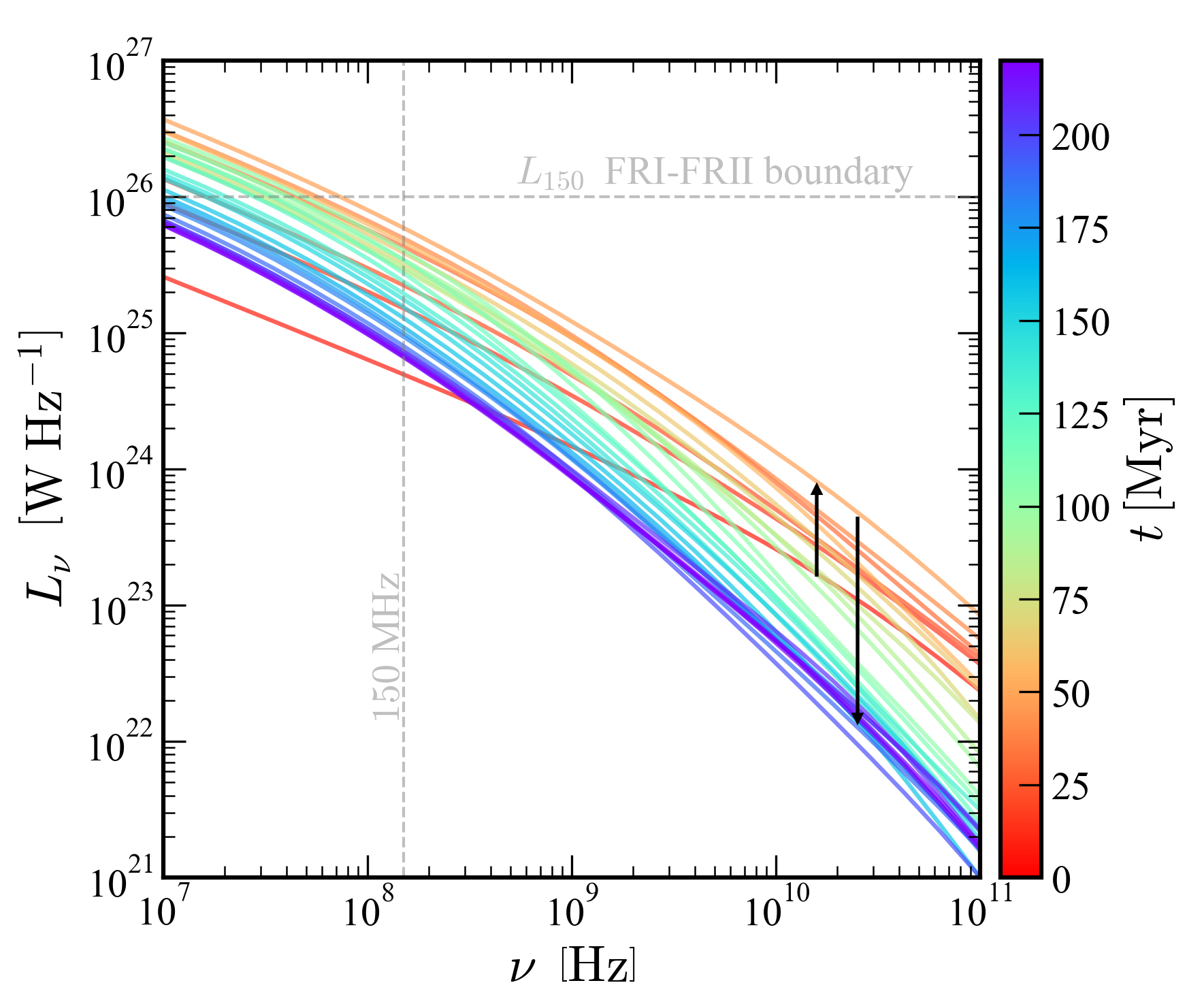}
	\caption{Time evolution of the total radio luminosity for a single jet outburst between 10~MHz and 100~GHz. The vertical dashed line indicates the $\nu = 150$~MHz frequency, and the horizontal dashed line the FRI-FRII radio power divide observed at this frequency \citep{Mingo2019}. The upward arrow indicates the build up of the spectrum through the 50~Myr of jet activity, while the downward arrow shows the evolution following jet termination. The spectral evolution over time is analogous to that of the CRe spectrum (see Fig.~\ref{fig:cre_spectrum}) and exhibits an increased curvature owing to synchrotron and inverse Compton cooling.}
	\label{fig:emission_spectrum}
\end{figure}

\subsection{Radio emission}

\begin{figure*}[ht]
	\centering
	\includegraphics[width=2\columnwidth]{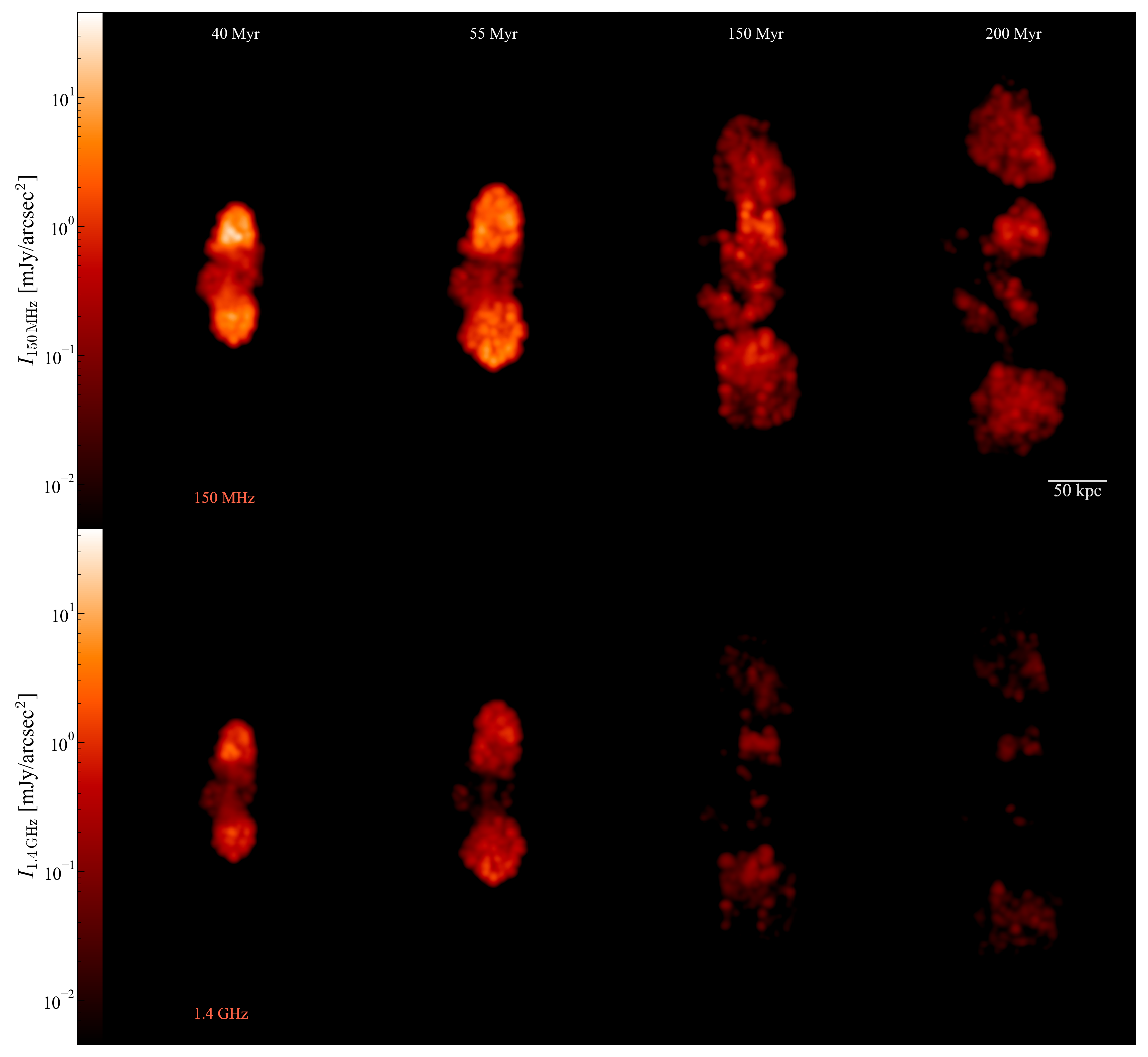}
	\caption{Synchrotron intensity maps of our single jet outburst at different epochs (left to right) at $150$~MHz and $1.4$~GHz (top and bottom). We integrated the synchrotron emissivity along a line of sight of $\pm 220$~kpc centred on the cluster. We smooth the intensity maps using a two-dimensional Gaussian beam with $10$~kpc full width half-maximum (FWHM). The high-frequency radio intensity is less spatially extended throughout the simulation due to faster cooling of high-energy electrons. This is especially visible at 200~Myr where the lobes have started fading out at $1.4$~GHz.}
	\label{fig:radio_maps}
\end{figure*}

We show radio emission spectra between 10~MHz and 100~GHz in Fig.~\ref{fig:emission_spectrum}. The temporal evolution is indicated using the colour bar and is analogous to the evolution of the CRe spectrum. The synchrotron spectrum builds up throughout jet activity (upwards arrow), after which synchrotron cooling dominates and causes the spectrum to decrease and curve, as expected from synchrotron theory. We show the FRI-FRII radio power divide at 150~MHz \citep{Fanaroff1974,Ledlow1996,Mingo2019}, indicated by a dashed line. Our jets produce radio powers consistent with observed powers of FRI radio galaxies.

In Fig.~\ref{fig:radio_maps}, we show synchrotron intensity maps at frequencies 150~MHz and 1.4~GHz at different epochs: 40, 55, 150 and 200~Myr. At 40~Myr, even though the jet is still active, the high frequency emission (bottom row) is less spatially extended (in the direction perpendicular to the jet axis) in comparison to the emission at 150~MHz. Intuitively, one might expect the high-frequency intensity to mirror the low-frequency emission when the jet is active and electrons are being continuously accelerated. However, these lateral regions where the high frequency emission is dimmer correspond to older electrons, as we will show in Sect.~\ref{sec:electron_ages}. These older electrons already exhibit the expected behaviour of faster synchrotron losses at higher frequencies, demonstrated in Fig.~\ref{fig:emission_spectrum}. This same effect is visible at 55~Myr. At 150~Myr and 200~Myr, electrons remain detectable at 150~MHz but have almost faded away at 1.4~GHz. This general feature, of the lower frequency radio intensity being more extended, is a consequence of electron populations having different ages and synchrotron cooling being faster at higher frequencies. This underscores the importance of employing spectral modelling to distinguish between different electron populations, accelerated and cooled over distinct timescales, as they are advected with the jet material into the ICM. At late times, the CRe populations in the wake of the lobes display intensities similar to those in the lobes, which is a consequence of converging flows compressing the central CRs (see also Fig.~\ref{fig:app_crep_projections}). The patchiness observed in the intensity maps, particularly at late times, is due to the tracer number staying constant once the jet shuts off (tracer particles are created in the jet, see Sect.~\ref{subsubsec:cre_acceleration}), causing the tracer number density  to decrease with time as the lobes rise and expand into the ICM.

\subsection{Connecting cosmic ray electron spectra and emission}\label{sec:connecting-cre-spectra-and-emission}

\begin{figure*}
	\centering
	\includegraphics[width=2\columnwidth]{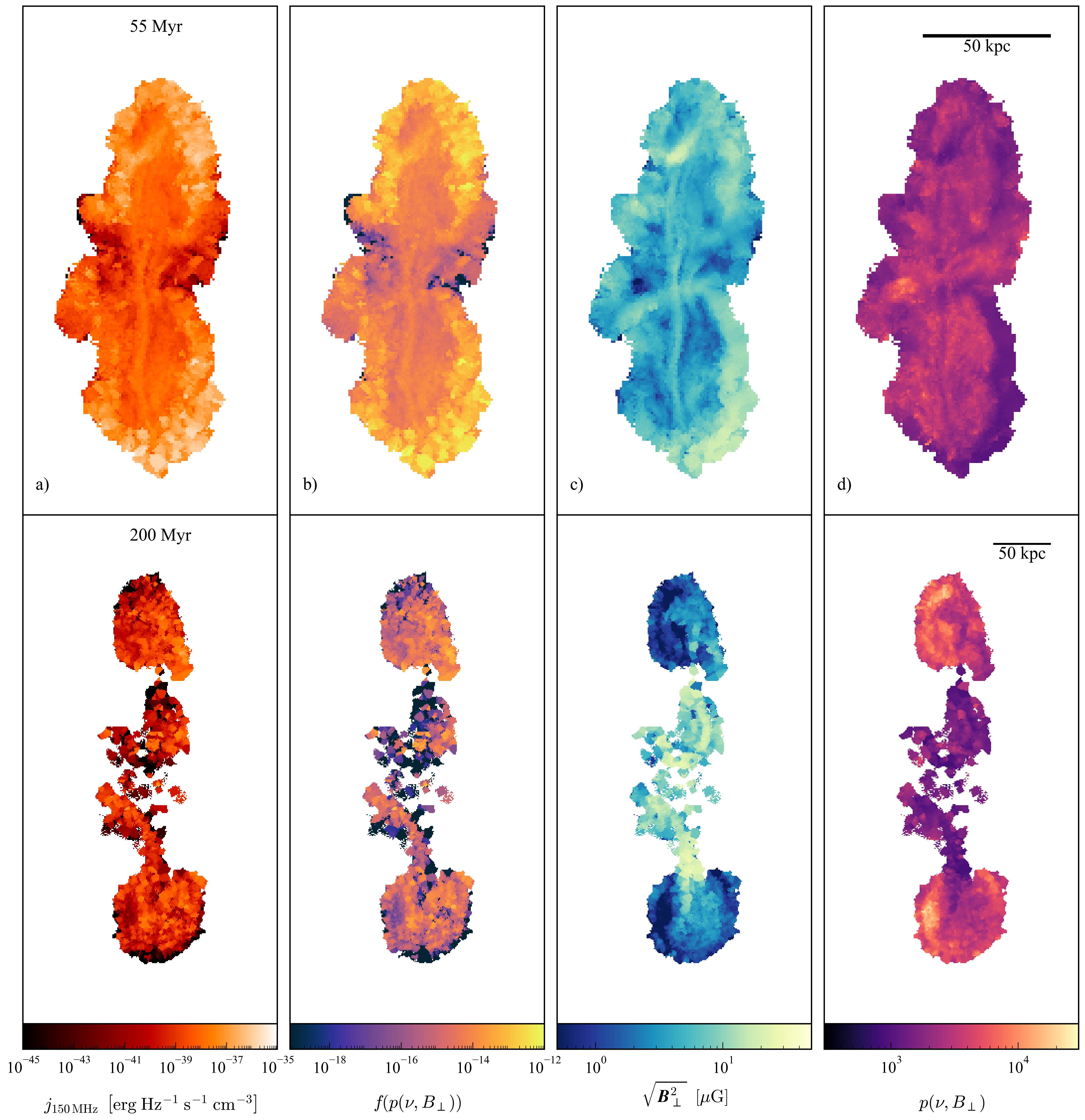}
	\caption{Thin projections of depth $\pm~7$~kpc shown at two different times for the following volume-averaged quantities, from left to right: \textit{a)} synchrotron emissivity at $\nu = 150$~MHz, \textit{b)} value of the distribution function at $p(\nu, B_{\perp})$, \textit{c)} magnetic field perpendicular to the line of sight, \textit{d)} the momentum which contributes most to the emissivity in panel \textit{a)}. 
	We recover the effect expected from equation~\eqref{eq:nu_sync}, whereby electrons with smaller momenta contribute most to the emissivity at a given frequency in regions of strong magnetic fields. A movie can be viewed \href{https://www.youtube.com/watch?v=hqPSUHKKsbs}{here}.}
	\label{fig:emission_slices}
\end{figure*}

Given the electron spectral information, we would like to understand which electron momentum dominates the synchrotron emission at a given frequency. To do so, we show thin projections of depth $\pm$7~kpc of different quantities related to CRe spectra and emission in Fig.~\ref{fig:emission_slices}. In panel~a), we  show the synchrotron emissivity at $\nu_{\rmn{sync}}=150$~MHz. In panel~b), we show the value of the distribution function $f(p)$ at the contributing momenta $p(\nu, B_{\perp})$. In panel~c), we show the magnetic field perpendicular to the line of sight which determines the contributing momentum $p(\nu, B_{\perp})$. In panel~d), we show the momenta $p(\nu, B_{\perp})$ that contribute most to the emission in panel~a) given the magnetic field perpendicular to the line of sight shown in panel~c). We remind the reader of the relationship between contributing momentum for a given frequency, $p^2 \propto \nu_{\rmn{sync}}/B_{\perp}$ (cf. Eq.~\ref{eq:nu_sync}) which is governed by the magnetic field perpendicular to the line of sight (see Sect.~\ref{subsec:crayon_emission}). Thus, in the presence of stronger magnetic fields, electrons with smaller momenta emit in a given frequency window and vice-versa. We call this the `$\nu_{\rm{c}}$-effect' \citep{Lacki2010, Werhahn2021c}.

At $t=55 \,\rmn{Myr}$ (upper panels), right after jet activity, the synchrotron emissivity traces the underlying magnetic field structure. The distribution of contributing momenta, within the range $10^3 \leq p \leq 10^{4}$, is a consequence of the varied magnetic field strengths. In regions of high magnetic fields such as in the central 50~kpc where a filament is visible, the momenta responsible for the $150\, \rmn{MHz}$ emission are low. On the other hand, the low magnetic fields in the lobe regions connected to the magnetic filaments lead to higher momenta being responsible for the observed emission. The regions of low emissivity and $f(p)$ observed in the regions perpendicular to the jet direction correspond to older populations, which we will discuss in Sect.~\ref{sec:electron_ages}. At late times, at $t=200 \,\rmn{Myr}$ (lower panels), after the jets have stopped, the synchrotron emissivity has decreased on average and has a more complex morphology in the wake of the lobes. Specifically, the emissivity and corresponding $f(p)$ are smoother in the lobes than in their tails, where the distribution is patchy. We see regions of weak magnetic fields in the lobes due to their expansion in the cluster atmosphere, while fields in the wake of the bubbles have been amplified. Correspondingly, these regions contain higher and lower contributing momenta, respectively, because of the $\nu_{\rm{c}}$-effect. The range of momenta at this epoch is within $10^3 \leq p \leq 3\times10^{4}$ and thus reaches slightly higher momenta than at early times. It can be seen that these regions with the highest momenta are also those with the lowest magnetic field in the southern lobe. To explain the patchiness of the emissivity and distribution function seen in panels a) and b) at late times, we now turn to analysing electron ages.

\subsection{Electron ages}\label{sec:electron_ages}

\begin{figure}
\centering
    \includegraphics[width=0.8\columnwidth]{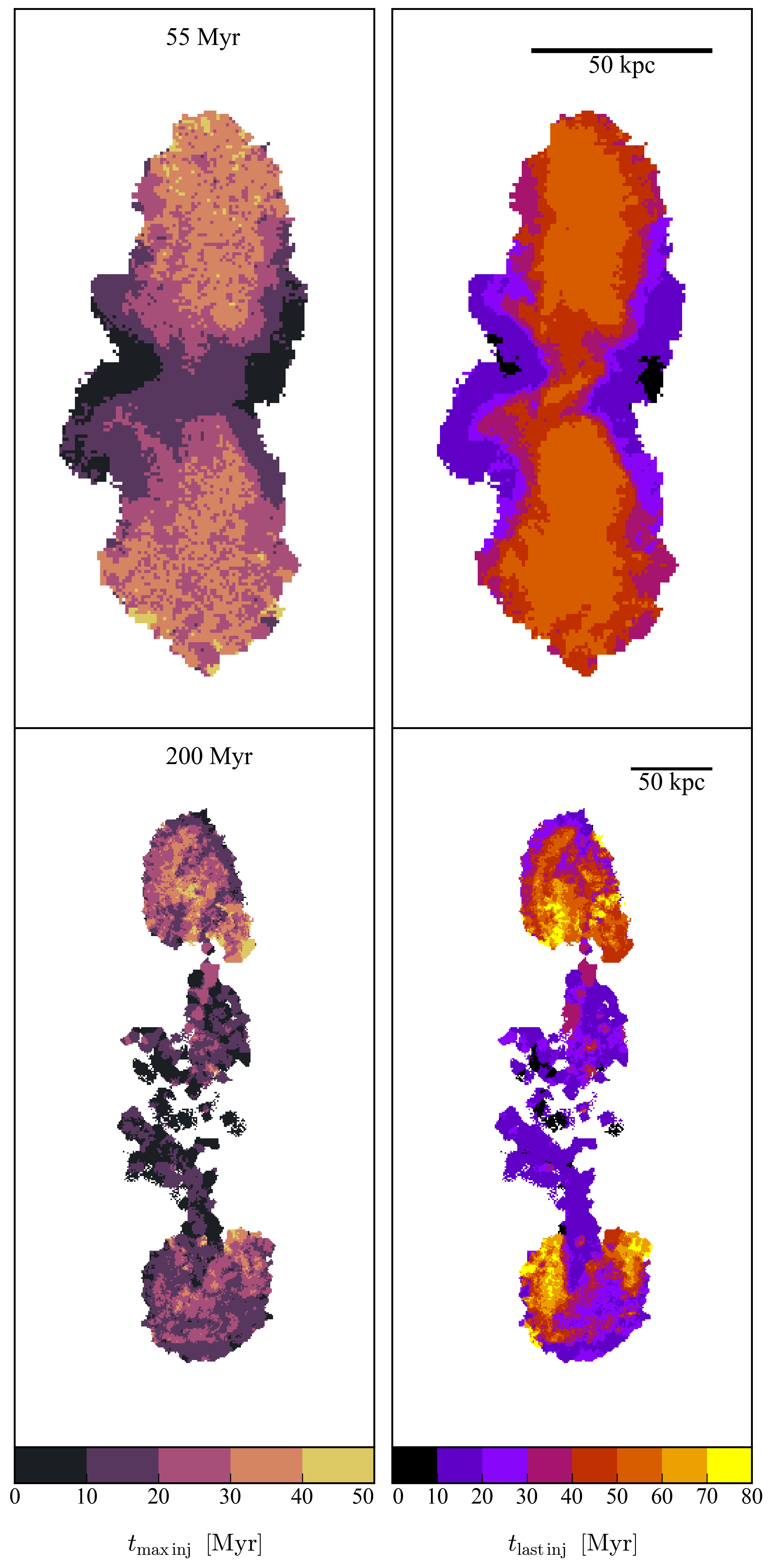}
    \caption{Thin volume-averaged projections of depth $\pm 7$~kpc. \textit{Left:} time of maximum injection, when the injected energy density is highest. \textit{Right:} time of last injection, when the cumulative injected energy density of a Lagrangian tracer has reached 99.7\% of its final value.
    Shortly after the jet switches off, at 55~Myr (upper panels), there is a clear gradient from younger electrons located around the jet spine towards older electrons that are situated at the edges of the lobes, perpendicular to the jet direction. Post jet activity, this trend is not as clear, but the uplift of older electrons is noticeable in the turbulent wake of the lobes. A movie can be viewed \href{https://www.youtube.com/watch?v=9BAirX3n4PA}{here}.}
    \label{fig:tinj_slices}
\end{figure}

Going beyond connecting emissivity at a given frequency to a given momentum, we can now attempt to define and identify electron ages, in order to understand which populations are contributing to the emission in Fig.~\ref{fig:emission_slices}. Looking at the idealized spectra for an exponentially decreasing source function in Fig.~\ref{fig:onezone}, it is clear that the electron age depends on the momentum of interest. For momenta $10 \leq p \leq 10^{4}$, where the cooling times are long, the contributing injection event is the one with the maximum normalization, i.e. with the maximum injected energy: the age of this specific electron population is thus the time of maximum injection $t_{\rmn{max\,inj}}$. Momenta $p > 10^{4}$ are within the steady-state region where cooling is fast: electron ages in this range are given by the time of the last injection event $t_{\rmn{last\,inj}}$. The time of last injection corresponds to the time at which the last CRe acceleration event occurred for a given tracer according to the criteria described in Sect.~\ref{subsubsec:cre_acceleration}. We limit $t_{\rmn{last\,inj}}$ to the time when $99.7\%$ ($3 \sigma$) of the cumulative injected electron energy has been reached for a given tracer. This is done to exclude acceleration events where a negligible fraction of the total energy is injected, which is a consequence of our progressive, exponentially decreasing acceleration algorithm.

The transition momentum (discussed in Sect.~\ref{sec:electron-spectra}) -- the momentum which separates the non-cooled and the steady-state parts of the spectrum -- is also the momentum which distinguishes these two CRe age definitions. For a given particle spectrum, however, the transition depends on the magnetic field strengths encountered by each particle along its trajectory. Stronger magnetic fields cause the transition momentum to shift to smaller momenta as CRes experience faster synchrotron losses -- momenta $p > 10^4$ cool within $\sim$3~Myr for CRes in a $30\, \mu \rm{G}$ field (cf.\ Fig.~\ref{fig:app_cooling_times}). On the other hand, tracer trajectories along weaker magnetic fields display transition momenta at higher momenta -- momenta $p > 10^4$ instead cool within 200~Myr in a CMB-equivalent magnetic field of $3.2\, \mu \rm{G}$. This transition momentum thus varies across particle spectra. For this reason, the age of the CRe population responsible for the observed synchrotron emission at a given time depends both on the history of the electrons -- which has shaped where the transition momentum lies -- and the current magnetic field -- which determines which part of the spectrum dominates the emission.

In Fig.~\ref{fig:tinj_slices}, we show thin projections of depth $\pm$7~kpc of the volume-weighted time of maximum injection, $t_{\rmn{max\,inj}}$, and time of last injection, $t_{\rmn{last\,inj}}$. A few Myr after the jet-active phase, the age gradient of the electron populations is noticeable in both maps. Focusing first on $t_{\rmn{max\,inj}}$, we find the youngest electrons (orange, $30 \leq t_{\rmn{max\,inj}}/\rmn{Myr} \leq 40$) embedded within slightly older electrons (pink, $20 \leq t_{\rmn{max\,inj}}/\rmn{Myr} \leq 30$), which are subsequently embedded in older and older electrons (purple and black). The oldest electrons, with $0 \leq t_{\rmn{max\,inj}}/\rmn{Myr} \leq 10$ (black), are visible perpendicular to the jet direction, at the furthest emitting edges. They have experienced more cooling, which explains the low synchrotron emissivity seen in these regions in Fig.~\ref{fig:emission_slices}. This gradient of increasing age from the jet spine/lobe towards the edge of the lobes is also observed in the $t_{\rmn{last\,inj}}$ map at early times. Furthermore, we see the effect of the progressive acceleration by focusing our attention on the central 50~kpc: while electrons have initially been accelerated at $\sim$10--20~Myr (purple, $10 \leq t_{\rmn{max\,inj}} \leq 20$), they are still experiencing acceleration at the current time of 55~Myr, as demonstrated by the right-hand panel, where $30 \leq t_{\rmn{last\,inj}}/\rmn{Myr} \leq 50$ (magenta to red) in the same central region. At later times when the lobes have risen and expanded in the cluster atmosphere, the age gradient is no longer as discernible. Although the two lobes display different $t_{\rmn{last\,inj}}$ spatial distributions, both lobes have funnelled older electrons due to a vortex forming (see \href{https://www.youtube.com/watch?v=9BAirX3n4PA}{movie}), an effect also studied by \citet{Chen2023}. This is noticeable in Fig.~\ref{fig:tinj_slices} in the northern lobe through old electrons lining the edge of the lobe (purple edge, $10 \leq t_{\rmn{last\,inj}}/\rmn{Myr} \leq 30$). In the southern lobe, older electrons are in the central part of the lobe and younger ones in the edges; the vortex funnelling older electrons upwards from the bubble wake is still ongoing.

We can now compare the distribution of ages seen in Fig.~\ref{fig:tinj_slices} and the values of the contributing momenta in Fig.~\ref{fig:emission_slices}. At 200~Myr, the electrons in the central 100~kpc are the oldest, with $0 \leq t_{\rmn{max\,inj}}/\rmn{Myr} \leq 20$ and $10 \leq t_{\rmn{last\,inj}}/\rmn{Myr} \leq 30$. This region corresponds to high magnetic fields as seen in panel c) of Fig.~\ref{fig:emission_slices}, where smaller electron momenta dominate the observed emission. The age of these low momenta electrons is thus best described by the time of maximum injection. Additionally, in the southern lobe, regions of high contributing momenta between $3\times10^{3} \leq p \leq 2\times10^{4}$, whose age is best described by the time of last injection, are spatially coincident with regions dominated by the youngest electrons (red-yellow, $60 \leq t_{\rmn{last\,inj}}/\rmn{Myr} \leq 80$). In the next step, we combine the previous sections by connecting electron spectra, ages and their emission. 

\begin{figure*}[!htb]
    \centering
    \includegraphics[width=1.6\columnwidth]{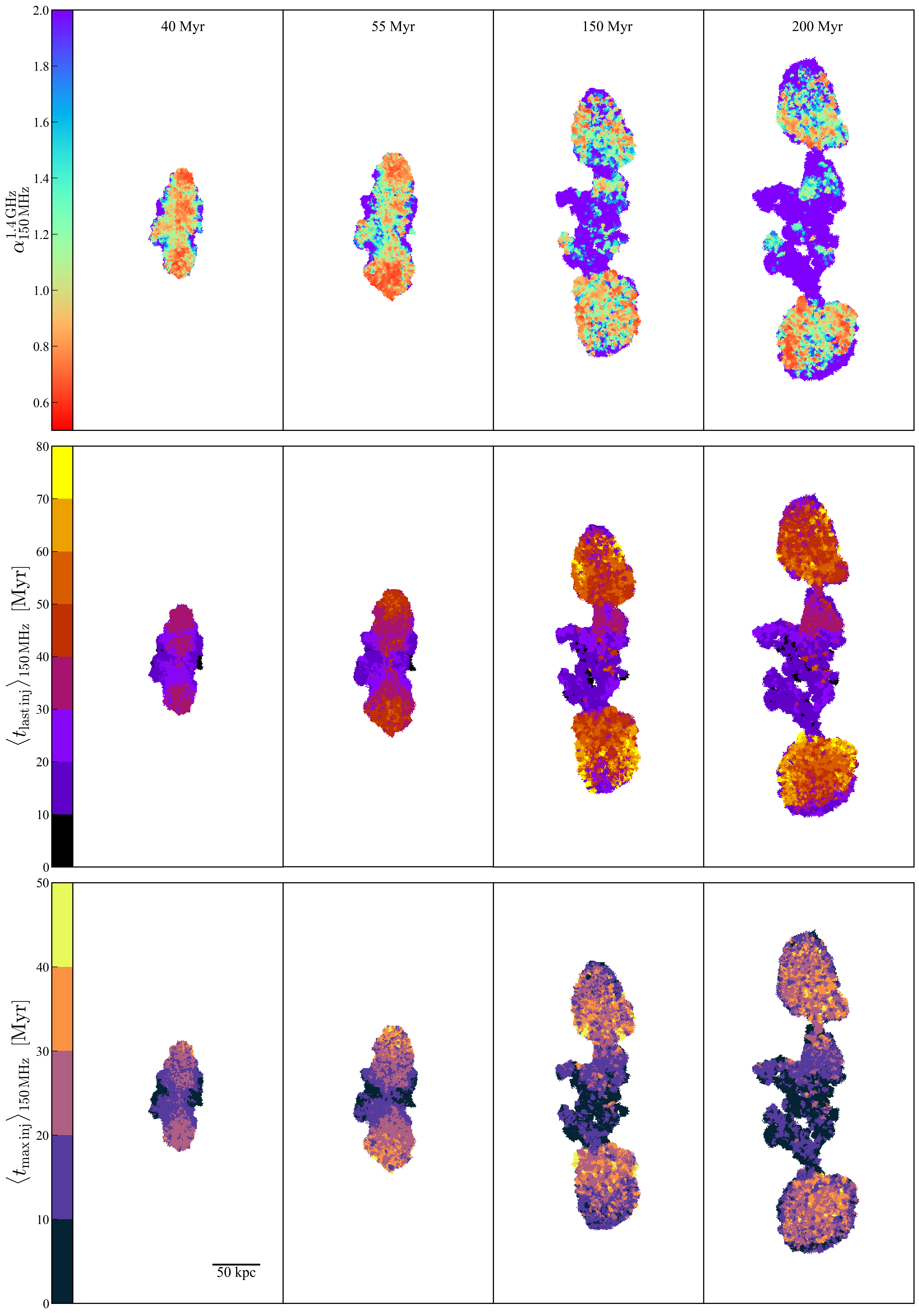}
    \caption{
    \textit{Top:} multi-epoch spectral index maps between $150$~MHz and $1.4$~GHz, where larger values correspond to older, more cooled plasma.
    \textit{Middle and bottom:} thick projections of depth $\pm 220$~kpc centred on the cluster showing the time of last injection and the time of maximum injection, respectively, weighted by the synchrotron luminosity at $150$~MHz. Different populations of electrons contribute along the line of sight, depending on the local magnetic field strength perpendicular to the line of sight. While the low spectral indices observed at early times are explained by ongoing CR acceleration, those observed at late times are due to adiabatic compression, which shifts the slower cooling part of the spectrum towards larger frequencies.}
    \label{fig:spectral_index_maps}
\end{figure*}

\subsection{Spectral index maps}

We create spectral index maps using Eq.~\eqref{eq:spectral_index} between 150~MHz and 1.4~GHz in the top row of Fig.~\ref{fig:spectral_index_maps}. We additionally show deep projections ($\pm$~220~kpc) of $t_{\rmn{max\,inj}}$ and $t_{\rmn{last\,inj}}$ weighted by the luminosity at 150~MHz, which allows us to compare spectral indices with electron ages. We expect radio spectral indices at around $0.6$ to correspond to freshly injected electrons, given the synchrotron spectral index for a population of electrons with injection index of $\alpha_\rmn{{inj}} = 2.2$. Larger spectral indices (green and blue), on the other hand, trace older populations of electrons, cooled through synchrotron and inverse Compton processes. On the bottom rows showing times of injection, purple values correspond to older electron populations, and yellow values correspond to younger electrons.
		
During jet activity ($t = 40$~Myr), electrons in the jet are experiencing acceleration and display a radio spectral index of $\sim$0.6. At 55~Myr, after the jet has switched off, some regions still display spectral indices of 0.6, due to the progressive, exponential nature of electron acceleration in our work. As the emitting electrons cool and rise in the cluster atmosphere, the spectral index distribution becomes patchy and varies from point to point (at 150~Myr), albeit with higher spectral indices seen in the tails of the bubbles.
At 150~Myr and 200~Myr, CRes in the wake of the bubbles corresponding to the oldest electrons have significantly aged, displaying spectral indices larger than 2. The lobes, on the other hand, show a large range of spectral values. Specifically, low spectral indices are observed in the edge of the southern bubble, close to $\sim$0.6. Naively, one would expect this to correspond to freshly accelerated electrons. Although these regions are coincident with younger electrons, as shown by the yellow colour in the middle panels showing $t_{\rmn{last\,inj}}$, no CRe populations are accelerated after $t=80$~Myr. Upon further investigation, we found that these low spectral indices are not observed at late times in our \textsc{Crest} model without adiabatic changes and mixing. This suggests that adiabatic compression shifts electrons to higher momenta (see left-hand panel of Fig.~\ref{fig:onezone_adiabatic}), where they emit into the 150~MHz -- 1.4 GHz radio window, thereby explaining these low spectral indices.

At 200~Myr, there appears to be a gradient within the lobes, where larger spectral indices are located in the lobe centres and lower indices at the lobes edges. This is noticeable by looking at the time of last injection in the southern lobe, which also shows this gradient: old electrons are located in the centre, and younger electrons are closer to the lobes edges. This is due to a vortex formation, whereby older electrons from the wake of the bubble are funnelled upwards through the lobe (discussed in \ref{sec:electron_ages}).

\begin{figure}[t]
	\centering
	\includegraphics[width=1.\columnwidth]{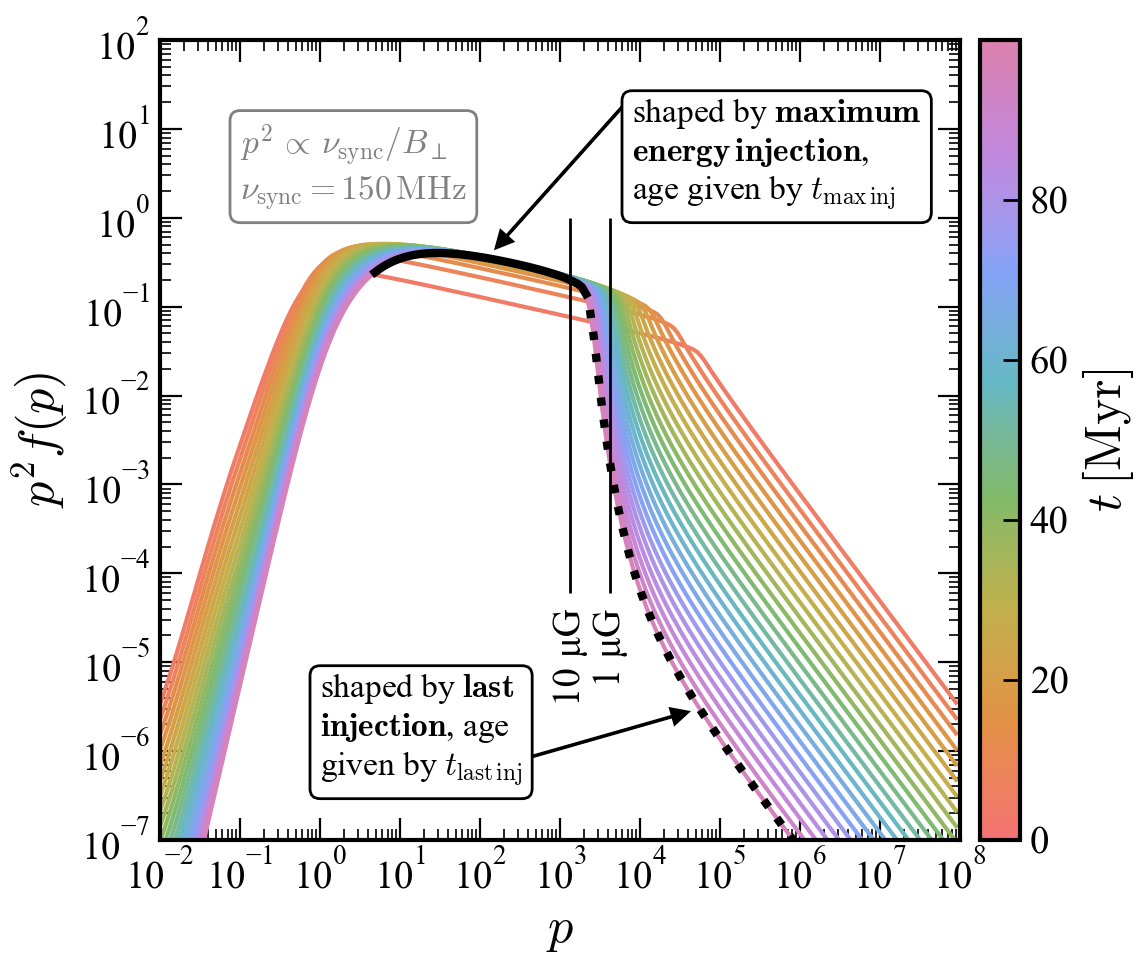}
	\caption{Figure summarizing which age definition connects to which momentum range in the electron spectrum depending on the magnetic field strength. For a tracer spectrum at a given time, the magnetic field strength determines which momentum dominates the emission at a given frequency (here $\nu_{\rm{sync}} = 150$~MHz) according to the $\nu_{\rmn{c}}$-effect. Lower momenta (solid black line) dominate the emission in strong magnetic fields (here $B=10 \, \mu \rmn{G}$), where the age of the spectrum is given by the time of maximum injection energy, $t_{\rmn{max\,inj}}$. Higher momenta (dotted black line) dominate the emission in weak magnetic fields (here $B=1 \, \mu \rmn{G}$), where the age of the spectrum is given by the time of last injection, $t_{\rmn{last\,inj}}$. We note that this CRe spectrum is an idealized spectrum. In our simulation, each particle spectrum is unique and shaped by its own history.}
	\label{fig:summary}
\end{figure}

At 200~Myr, looking at the northern lobe and the uplifted electrons in the central 100~kpc, spectral ages do not evidently favour either $t_{\rmn{max\,inj}}$ or $t_{\rmn{last\,inj}}$ in explaining the observed structure of ages. Indeed, recalling the $\nu_{\rm{c}}$-effect, the momentum which contributes most to the emission at a given frequency depends on the value of the magnetic field perpendicular to the line of sight: lower momenta dominate in stronger magnetic fields and vice-versa, as summarized in Fig.~\ref{fig:summary}. This means that the higher the magnetic field, the lower the momentum that dominates the emission: the electron age is then best described by the time of maximum injection. In lower magnetic fields, higher momenta dominate the emission: the electron age is therefore best described by the last injection time. Within the same lobe at a given time and along each line of sight, different populations are contributing to the emission (and hence the spectral age) depending on the magnetic field perpendicular to the line of sight of a given particle.

\section{Conclusions}\label{sec:conclusions}
We have performed MHD simulations of a single AGN-jet outburst in an idealized Perseus-like cool core galaxy cluster and presented the evolution of non-thermal electron populations. This has enabled us to test our sub-grid acceleration algorithms, how they influence the production of synchrotron emission and to study the link between electron ages and spectral indices. We have developed sub-grid algorithms for acceleration of CRps and CRes in which a fraction of the jet energy is progressively injected over time. We evolve CRe spectra using the \textsc{Crest} Fokker-Planck solver, and produce non-thermal emission spectra and maps using \textsc{Crayon+}. We summarize our results in the following:

\begin{itemize}
	\item Our CR acceleration models correspond to an exponentially decreasing source function across time. The resulting electron spectrum resembles a freely-cooling spectrum at low- and mid-momenta and a steady-state spectrum at high momenta with a decreasing normalization of the high-momentum power law (Figs.~\ref{fig:onezone} and \ref{fig:cre_spectrum}).
    
    \item Our approach allows us to connect electron spectra to the resulting synchrotron emission using the background MHD properties (Fig.~\ref{fig:emission_slices}). According to the $\nu_{\rmn{c}}$-effect, a given emission frequency is dominated by lower (higher) momenta in stronger (weaker) magnetic fields. These regimes thus impact which electron age is attributed to the observed emission at a given frequency. Lower momenta dominate the emission in strong magnetic fields -- the electron age is best determined by the time of maximum injected energy. Higher momenta dominate the emission in weak magnetic fields -- the electron age is best described by the time of last injection (Fig.~\ref{fig:summary}).
    
    \item The total electron spectrum at momenta $p > 10^2$ is mostly unchanged when adiabatic changes and mixing effects are included (Fig.~\ref{fig:cre_spectrum}), although there are local variations in individual CRe populations (Fig.~\ref{fig:spectral_index_maps}). This highlights that CRs accelerated in low-density jets in a cool-core cluster environment are simultaneously compressed (shift of the spectrum right- and upwards) and diluted (causing a downward shift of the spectrum) in such a way that the total normalization remains nearly invariant. This is due to low-density jet plasma mixing with the dense ICM.
    
\end{itemize}

The combination of codes (\textsc{Arepo}, \textsc{Crest}, \textsc{Crayon+}) and algorithms (AGN jet feedback, CRp and CRe acceleration models) used in this work allows us to accelerate, advect and cool CRe on large time and spatial scales. Especially in systems such as cool-core clusters where AGN feedback operates, the central radio galaxy is thought to be always on \citep{Sabater2019}. Simulating the evolution of non-thermal electrons throughout multiple jet events, which regulate cooling flows and inject turbulence into the ambient ICM, will be a crucial step to produce the complex morphologies observed in radio galaxies. We will tackle this in future work.

\begin{acknowledgements}
   The analysis of the simulations presented in this work were performed using the Python package \textsc{Paicos} \citep{Berlok2024}. LJ and CP  acknowledge support from the Deutsche Forschungsgemeinschaft (DFG, German Research Foundation) as part of the DFG Research Unit FOR5195 – project number 443220636. CP and JW acknowledge support by the European Research Council under ERC-AdG grant PICOGAL-101019746. RW acknowledges funding of a Leibniz Junior Research Group (project number J131/2022). JW acknowledges support by the German Science Foundation (DFG) under grant 444932369. PG gratefully acknowledges financial support from the European Research Council via the ERC Synergy Grant "ECOGAL" (project ID 855130).
\end{acknowledgements}

\section*{Data Availability}

The data underlying this article will be shared on reasonable request to the corresponding author.
%



\bibliographystyle{aa}
\bibliography{paper1.bib}

@article{Aharonian2010,
  title = {Angular, Spectral, and Time Distributions of Highest Energy Protons and Associated Secondary Gamma-Rays and Neutrinos Propagating through Extragalactic Magnetic and Radiation Fields},
  author = {Aharonian, F. A. and Kelner, S. R. and Prosekin, A. Y.},
  year = 2010,
  month = aug,
  journal = {Physical Review D},
  volume = {82},
  number = {4},
  eprint = {1006.1045},
  primaryclass = {astro-ph},
  pages = {043002},
  issn = {1550-7998, 1550-2368},
  doi = {10.1103/PhysRevD.82.043002},
  urldate = {2025-03-10},
  archiveprefix = {arXiv},
  langid = {english}
}

@inproceedings{Axford1977,
  title = {The Acceleration of Cosmic Rays by Shock Waves},
  booktitle = {International {{Cosmic Ray Conference}}},
  author = {Axford, W. I. and Leer, E. and Skadron, G.},
  year = 1977,
  month = jan,
  volume = {11},
  pages = {132},
  urldate = {2025-04-15}
}

@article{Beckmann2022,
  title = {Cosmic Rays and Thermal Instability in Self-Regulating Cooling Flows of Massive Galaxy Clusters},
  author = {Beckmann, Ricarda S. and Dubois, Yohan and Pellissier, Alisson and Olivares, Valeria and Polles, Fiorella L. and Hahn, Oliver and Guillard, Pierre and Lehnert, Matthew D.},
  year = 2022,
  month = sep,
  journal = {Astronomy \& Astrophysics},
  volume = {665},
  pages = {A129},
  issn = {0004-6361, 1432-0746},
  doi = {10.1051/0004-6361/202142527},
  urldate = {2025-06-19},
  copyright = {https://creativecommons.org/licenses/by/4.0},
  langid = {english}
}

@article{Bell1978,
  title = {The Acceleration of Cosmic Rays in Shock Fronts -- {{II}}},
  author = {Bell, A. R.},
  year = 1978,
  month = mar,
  journal = {Monthly Notices of the Royal Astronomical Society},
  volume = {182},
  number = {3},
  pages = {443--455},
  issn = {0035-8711},
  doi = {10.1093/mnras/182.3.443},
  urldate = {2025-04-15}
}

@article{Bell1978a,
  title = {The Acceleration of Cosmic Rays in Shock Fronts - {{I}}.},
  author = {Bell, A. R.},
  year = 1978,
  month = jan,
  journal = {Monthly Notices of the Royal Astronomical Society},
  volume = {182},
  pages = {147--156},
  publisher = {OUP},
  issn = {0035-8711},
  doi = {10.1093/mnras/182.2.147},
  urldate = {2025-04-15}
}

@article{Berlok2024,
  title = {Paicos: A Python Package for Analysis of (Cosmological)Simulations Performed with Arepo},
  shorttitle = {Paicos},
  author = {Berlok, Thomas and Jlassi, L{\'e}na and Puchwein, Ewald and Haugb{\o}lle, Troels},
  year = 2024,
  month = apr,
  journal = {Journal of Open Source Software},
  volume = {9},
  number = {96},
  pages = {6296},
  issn = {2475-9066},
  doi = {10.21105/joss.06296},
  urldate = {2024-04-22},
  copyright = {http://creativecommons.org/licenses/by/4.0/},
  langid = {english}
}

@article{Birzan2004,
  title = {A Systematic Study of Radio-Induced x-Ray Cavities in Clusters, Groups, and Galaxies},
  author = {Birzan, L. and Rafferty, D. A. and McNamara, B. R. and Wise, M. W. and Nulsen, P. E. J.},
  year = 2004,
  month = jun,
  journal = {The Astrophysical Journal},
  volume = {607},
  number = {2},
  eprint = {astro-ph/0402348},
  pages = {800--809},
  issn = {0004-637X, 1538-4357},
  doi = {10.1086/383519},
  urldate = {2023-04-11},
  archiveprefix = {arXiv},
  langid = {english}
}

@article{Birzan2008,
  title = {Radiative Efficiency and Content of Extragalactic Radio Sources: {{Toward}} a Universal Scaling Relation between Jet Power and Radio Power},
  shorttitle = {Radiative Efficiency and Content of Extragalactic Radio Sources},
  author = {B{\^i}rzan, L. and McNamara, B. R. and Nulsen, P. E. J. and Carilli, C. L. and Wise, M. W.},
  year = 2008,
  month = oct,
  journal = {The Astrophysical Journal},
  volume = {686},
  number = {2},
  pages = {859--880},
  issn = {0004-637X, 1538-4357},
  doi = {10.1086/591416},
  urldate = {2025-06-18},
  langid = {english}
}

@article{Blandford2019,
  title = {Relativistic Jets in Active Galactic Nuclei},
  author = {Blandford, Roger and Meier, David and Readhead, Anthony},
  year = 2019,
  month = aug,
  journal = {Annual Review of Astronomy and Astrophysics},
  volume = {57},
  number = {1},
  eprint = {1812.06025},
  primaryclass = {astro-ph},
  pages = {467--509},
  issn = {0066-4146, 1545-4282},
  doi = {10.1146/annurev-astro-081817-051948},
  urldate = {2023-04-11},
  archiveprefix = {arXiv},
  langid = {english}
}

@article{Brienza2017,
  title = {Search and Modelling of Remnant Radio Galaxies in the {{LOFAR}} Lockman Hole Field},
  author = {Brienza, M. and Godfrey, L. and Morganti, R. and Prandoni, I. and Harwood, J. and Mahony, E. K. and Hardcastle, M. J. and Murgia, M. and R{\"o}ttgering, H. J. A. and Shimwell, T. W. and Shulevski, A.},
  year = 2017,
  month = oct,
  journal = {Astronomy \& Astrophysics},
  volume = {606},
  pages = {A98},
  issn = {0004-6361, 1432-0746},
  doi = {10.1051/0004-6361/201730932},
  urldate = {2025-02-25},
  langid = {english}
}

@article{Brunetti2014,
  title = {Cosmic Rays in Galaxy Clusters and Their Non-Thermal Emission},
  author = {Brunetti, G. and Jones, T. W.},
  year = 2014,
  month = apr,
  journal = {International Journal of Modern Physics D},
  volume = {23},
  number = {04},
  eprint = {1401.7519},
  primaryclass = {astro-ph},
  pages = {1430007},
  issn = {0218-2718, 1793-6594},
  doi = {10.1142/S0218271814300079},
  urldate = {2023-04-11},
  archiveprefix = {arXiv},
  langid = {english}
}

@misc{Caprioli2019,
  title = {The {{Issue}} with {{Diffusive Shock Acceleration}}},
  author = {Caprioli, Damiano and Haggerty, Colby C.},
  year = 2019,
  month = sep,
  number = {arXiv:1909.06288},
  eprint = {1909.06288},
  primaryclass = {astro-ph},
  publisher = {arXiv},
  doi = {10.48550/arXiv.1909.06288},
  urldate = {2025-11-25},
  archiveprefix = {arXiv},
  langid = {english}
}

@article{Cerutti2023,
  title = {Extreme Ion Acceleration at Extragalactic Jet Termination Shocks},
  author = {Cerutti, Beno{\^i}t and Giacinti, Gwenael},
  year = 2023,
  month = aug,
  journal = {Astronomy \& Astrophysics},
  volume = {676},
  pages = {A23},
  issn = {0004-6361, 1432-0746},
  doi = {10.1051/0004-6361/202346481},
  urldate = {2025-07-01},
  copyright = {https://creativecommons.org/licenses/by/4.0},
  langid = {english}
}

@article{Chen2023,
  title = {A Numerical Study of the Impact of Jet Magnetic Topology on Radio Galaxy Evolution},
  author = {Chen, Yi-Hao and Heinz, Sebastian and Hooper, Eric},
  year = 2023,
  month = jun,
  journal = {Monthly Notices of the Royal Astronomical Society},
  volume = {522},
  number = {2},
  pages = {2850--2868},
  issn = {0035-8711},
  doi = {10.1093/mnras/stad1074},
  urldate = {2025-06-19}
}

@article{Churazov2003,
  title = {{{XMM-newton}} Observations of the Perseus Cluster {{I}}: {{The}} Temperature and Surface Brightness Structure},
  shorttitle = {{{XMM-newton}} Observations of the Perseus Cluster {{I}}},
  author = {Churazov, E. and Forman, W. and Jones, C. and B{\"o}hringer, H.},
  year = 2003,
  month = jun,
  journal = {The Astrophysical Journal},
  volume = {590},
  number = {1},
  eprint = {astro-ph/0301482},
  pages = {225--237},
  issn = {0004-637X, 1538-4357},
  doi = {10.1086/374923},
  urldate = {2023-04-11},
  archiveprefix = {arXiv},
  langid = {english}
}

@article{Croston2018,
  title = {Particle Content, Radio-Galaxy Morphology, and Jet Power: All Radio-Loud {{AGN}} Are Not Equal},
  shorttitle = {Particle Content, Radio-Galaxy Morphology, and Jet Power},
  author = {Croston, J H and Ineson, J and Hardcastle, M J},
  year = 2018,
  month = may,
  journal = {Monthly Notices of the Royal Astronomical Society},
  volume = {476},
  number = {2},
  pages = {1614--1623},
  issn = {0035-8711, 1365-2966},
  doi = {10.1093/mnras/sty274},
  urldate = {2025-02-25},
  langid = {english}
}

@article{Dubey2023,
  title = {Particles in Relativistic {{MHD}} Jets. {{I}}. {{Role}} of Jet Dynamics in Particle Acceleration},
  author = {Dubey, Ravi Pratap and Fendt, Christian and Vaidya, Bhargav},
  year = 2023,
  month = jul,
  journal = {The Astrophysical Journal},
  volume = {952},
  number = {1},
  pages = {1},
  publisher = {The American Astronomical Society},
  issn = {0004-637X},
  doi = {10.3847/1538-4357/ace0bf},
  urldate = {2025-06-19},
  langid = {english}
}

@article{Dunn2004,
  title = {Particle Energies and Filling Fractions of Radio Bubbles in Cluster Cores},
  author = {Dunn, R. J. H. and Fabian, A. C.},
  year = 2004,
  month = dec,
  journal = {Monthly Notices of the Royal Astronomical Society},
  volume = {355},
  number = {3},
  pages = {862--873},
  issn = {1365-2966, 0035-8711},
  doi = {10.1111/j.1365-2966.2004.08365.x},
  urldate = {2025-07-04},
  langid = {english}
}

@article{Dursi2008,
  title = {Draping of Cluster Magnetic Fields over Bullets and Bubbles---Morphology and Dynamic Effects},
  author = {Dursi, L. J. and Pfrommer, C.},
  year = 2008,
  month = apr,
  journal = {The Astrophysical Journal},
  volume = {677},
  number = {2},
  pages = {993--1018},
  issn = {0004-637X, 1538-4357},
  doi = {10.1086/529371},
  urldate = {2025-03-05},
  langid = {english}
}

@article{Ehlert2018,
  title = {Simulations of the Dynamics of Magnetized Jets and Cosmic Rays in Galaxy Clusters},
  author = {Ehlert, K and Weinberger, R and Pfrommer, C and Pakmor, R and Springel, V},
  year = 2018,
  month = dec,
  journal = {Monthly Notices of the Royal Astronomical Society},
  volume = {481},
  number = {3},
  pages = {2878--2900},
  issn = {0035-8711, 1365-2966},
  doi = {10.1093/mnras/sty2397},
  urldate = {2023-04-11},
  langid = {english}
}

@article{Ehlert2023,
  title = {Self-Regulated {{AGN}} Feedback of Light Jets in Cool-Core Galaxy Clusters},
  author = {Ehlert, K and Weinberger, R and Pfrommer, C and Pakmor, R and Springel, V},
  year = 2023,
  month = jan,
  journal = {Monthly Notices of the Royal Astronomical Society},
  volume = {518},
  number = {3},
  pages = {4622--4645},
  issn = {0035-8711},
  doi = {10.1093/mnras/stac2860},
  urldate = {2025-06-19}
}

@article{English2016,
  title = {Numerical Modelling of the Lobes of Radio Galaxies in Cluster Environments -- {{III}}. {{Powerful}} Relativistic and Non-Relativistic Jets},
  author = {English, W. and Hardcastle, M. J. and Krause, M. G. H.},
  year = 2016,
  month = sep,
  journal = {Monthly Notices of the Royal Astronomical Society},
  volume = {461},
  number = {2},
  pages = {2025--2043},
  issn = {0035-8711},
  doi = {10.1093/mnras/stw1407},
  urldate = {2025-01-14}
}

@article{Ensslin2007,
  title = {Cosmic Ray Physics in Calculations of Cosmological Structure Formation},
  author = {En{\ss}lin, T. A. and Pfrommer, C. and Springel, V. and Jubelgas, M.},
  year = 2007,
  month = oct,
  journal = {Astronomy \& Astrophysics},
  volume = {473},
  number = {1},
  pages = {41--57},
  issn = {0004-6361, 1432-0746},
  doi = {10.1051/0004-6361:20065294},
  urldate = {2023-04-11},
  langid = {english}
}

@article{Ensslin2011,
  title = {Cosmic Ray Transport in Galaxy Clusters: Implications for Radio Halos, Gamma-Ray Signatures, and Cool Core Heating},
  shorttitle = {Cosmic Ray Transport in Galaxy Clusters},
  author = {Ensslin, Torsten A. and Pfrommer, Christoph and Miniati, Francesco and Subramanian, Kandaswamy},
  year = 2011,
  month = mar,
  journal = {Astronomy \& Astrophysics},
  volume = {527},
  eprint = {1008.4717},
  primaryclass = {astro-ph},
  pages = {A99},
  issn = {0004-6361, 1432-0746},
  doi = {10.1051/0004-6361/201015652},
  urldate = {2023-04-11},
  archiveprefix = {arXiv},
  langid = {english}
}

@article{Fanaroff1974,
  title = {The Morphology of Extragalactic Radio Sources of High and Low Luminosity},
  author = {Fanaroff, B. L. and Riley, J. M.},
  year = 1974,
  month = may,
  journal = {Monthly Notices of the Royal Astronomical Society},
  volume = {167},
  pages = {31P-36P},
  issn = {0035-8711},
  doi = {10.1093/mnras/167.1.31P},
  urldate = {2023-04-11}
}

@article{Fermi1949,
  title = {On the Origin of the Cosmic Radiation},
  author = {Fermi, {\relax ENRICO}},
  year = 1949,
  month = apr,
  journal = {Physical Review},
  volume = {75},
  number = {8},
  pages = {1169--1174},
  publisher = {American Physical Society},
  doi = {10.1103/PhysRev.75.1169},
  urldate = {2025-04-15}
}

@article{Gendron-Marsolais2020,
  title = {High-Resolution {{VLA}} Low Radio Frequency Observations of the Perseus Cluster: Radio Lobes, Mini-Halo and Bent-Jet Radio Galaxies},
  shorttitle = {High-Resolution {{VLA}} Low Radio Frequency Observations of the Perseus Cluster},
  author = {{Gendron-Marsolais}, Marie-Lou and {Hlavacek-Larrondo}, Julie and {van Weeren}, Reinout J. and Rudnick, Lawrence and Clarke, Tracy E. and Sebastian, Biny and Mroczkowski, Tony and Fabian, Andrew C. and Blundell, Katherine M. and Sheldahl, Evan and Nyland, Kristina and Sanders, Jeremy S. and Peters, Wendy M. and Intema, Huib T.},
  year = 2020,
  month = nov,
  journal = {Monthly Notices of the Royal Astronomical Society},
  volume = {499},
  number = {4},
  eprint = {2005.12298},
  primaryclass = {astro-ph},
  pages = {5791--5805},
  issn = {0035-8711, 1365-2966},
  doi = {10.1093/mnras/staa2003},
  urldate = {2023-04-11},
  archiveprefix = {arXiv},
  langid = {english}
}

@article{Genel2013,
  title = {Following the Flow: Tracer Particles in Astrophysical Fluid Simulations},
  shorttitle = {Following the Flow},
  author = {Genel, Shy and Vogelsberger, Mark and Nelson, Dylan and Sijacki, Debora and Springel, Volker and Hernquist, Lars},
  year = 2013,
  month = oct,
  journal = {Monthly Notices of the Royal Astronomical Society},
  volume = {435},
  number = {2},
  eprint = {1305.2195},
  primaryclass = {astro-ph, physics:physics},
  pages = {1426--1442},
  issn = {1365-2966, 0035-8711},
  doi = {10.1093/mnras/stt1383},
  urldate = {2023-04-11},
  archiveprefix = {arXiv},
  langid = {english}
}

@article{Guo2008,
  title = {Feedback Heating by Cosmic Rays in Clusters of Galaxies},
  author = {Guo, Fulai and Oh, S. Peng},
  year = 2008,
  month = feb,
  journal = {Monthly Notices of the Royal Astronomical Society},
  volume = {384},
  number = {1},
  pages = {251--266},
  issn = {0035-8711, 1365-2966},
  doi = {10.1111/j.1365-2966.2007.12692.x},
  urldate = {2025-06-20},
  langid = {english}
}

@article{Guo2011,
  title = {Cosmic-Ray-Dominated Agn Jets and the Formation of x-Ray Cavities in Galaxy Clusters},
  author = {Guo, Fulai and Mathews, William G.},
  year = 2011,
  month = feb,
  journal = {The Astrophysical Journal},
  volume = {728},
  number = {2},
  pages = {121},
  issn = {0004-637X, 1538-4357},
  doi = {10.1088/0004-637X/728/2/121},
  urldate = {2025-06-20},
  langid = {english}
}

@article{Hardcastle2013,
  title = {Numerical Modelling of the Lobes of Radio Galaxies in Cluster Environments},
  author = {Hardcastle, M. J. and Krause, M. G. H.},
  year = 2013,
  month = mar,
  journal = {Monthly Notices of the Royal Astronomical Society},
  volume = {430},
  number = {1},
  pages = {174--196},
  issn = {1365-2966, 0035-8711},
  doi = {10.1093/mnras/sts564},
  urldate = {2025-04-15},
  langid = {english}
}

@article{Hardcastle2014,
  title = {Numerical Modelling of the Lobes of Radio Galaxies in Cluster Environments -- {{II}}. {{Magnetic}} Field Configuration and Observability},
  author = {Hardcastle, M. J. and Krause, M. G. H.},
  year = 2014,
  month = sep,
  journal = {Monthly Notices of the Royal Astronomical Society},
  volume = {443},
  number = {2},
  pages = {1482--1499},
  issn = {0035-8711},
  doi = {10.1093/mnras/stu1229},
  urldate = {2025-01-14}
}

@article{HitomiCollaboration2018,
  title = {Atmospheric Gas Dynamics in the Perseus Cluster Observed with Hitomi},
  author = {{Hitomi Collaboration} and Aharonian, Felix and Akamatsu, Hiroki and Akimoto, Fumie and Allen, Steven W and Angelini, Lorella and Audard, Marc and Awaki, Hisamitsu and Axelsson, Magnus and Bamba, Aya and Bautz, Marshall W and Blandford, Roger and Brenneman, Laura W and Brown, Gregory V and Bulbul, Esra and Cackett, Edward M and Canning, Rebecca E A and Chernyakova, Maria and Chiao, Meng P and Coppi, Paolo S and Costantini, Elisa and De Plaa, Jelle and De Vries, Cor P and Den Herder, Jan-Willem and Done, Chris and Dotani, Tadayasu and Ebisawa, Ken and Eckart, Megan E and Enoto, Teruaki and Ezoe, Yuichiro and Fabian, Andrew C and Ferrigno, Carlo and Foster, Adam R and Fujimoto, Ryuichi and Fukazawa, Yasushi and Furuzawa, Akihiro and Galeazzi, Massimiliano and Gallo, Luigi C and Gandhi, Poshak and Giustini, Margherita and Goldwurm, Andrea and Gu, Liyi and Guainazzi, Matteo and Haba, Yoshito and Hagino, Kouichi and Hamaguchi, Kenji and Harrus, Ilana M and Hatsukade, Isamu and Hayashi, Katsuhiro and Hayashi, Takayuki and Hayashi, Tasuku and Hayashida, Kiyoshi and Hiraga, Junko S and Hornschemeier, Ann and Hoshino, Akio and Hughes, John P and Ichinohe, Yuto and Iizuka, Ryo and Inoue, Hajime and Inoue, Shota and Inoue, Yoshiyuki and Ishida, Manabu and Ishikawa, Kumi and Ishisaki, Yoshitaka and Iwai, Masachika and Kaastra, Jelle and Kallman, Tim and Kamae, Tsuneyoshi and Kataoka, Jun and Katsuda, Satoru and Kawai, Nobuyuki and Kelley, Richard L and Kilbourne, Caroline A and Kitaguchi, Takao and Kitamoto, Shunji and Kitayama, Tetsu and Kohmura, Takayoshi and Kokubun, Motohide and Koyama, Katsuji and Koyama, Shu and Kretschmar, Peter and Krimm, Hans A and Kubota, Aya and Kunieda, Hideyo and Laurent, Philippe and Lee, Shiu-Hang and Leutenegger, Maurice A and Limousin, Olivier and Loewenstein, Michael and Long, Knox S and Lumb, David and Madejski, Greg and Maeda, Yoshitomo and Maier, Daniel and Makishima, Kazuo and Markevitch, Maxim and Matsumoto, Hironori and Matsushita, Kyoko and McCammon, Dan and McNamara, Brian R and Mehdipour, Missagh and Miller, Eric D and Miller, Jon M and Mineshige, Shin and Mitsuda, Kazuhisa and Mitsuishi, Ikuyuki and Miyazawa, Takuya and Mizuno, Tsunefumi and Mori, Hideyuki and Mori, Koji and Mukai, Koji and Murakami, Hiroshi and Mushotzky, Richard F and Nakagawa, Takao and Nakajima, Hiroshi and Nakamori, Takeshi and Nakashima, Shinya and Nakazawa, Kazuhiro and Nobukawa, Kumiko K and Nobukawa, Masayoshi and Noda, Hirofumi and Odaka, Hirokazu and Ohashi, Takaya and Ohno, Masanori and Okajima, Takashi and Ota, Naomi and Ozaki, Masanobu and Paerels, Frits and Paltani, St{\'e}phane and Petre, Robert and Pinto, Ciro and Porter, Frederick S and Pottschmidt, Katja and Reynolds, Christopher S and {Safi-Harb}, Samar and Saito, Shinya and Sakai, Kazuhiro and Sasaki, Toru and Sato, Goro and Sato, Kosuke and Sato, Rie and Sawada, Makoto and Schartel, Norbert and Serlemtsos, Peter J and Seta, Hiromi and Shidatsu, Megumi and Simionescu, Aurora and Smith, Randall K and Soong, Yang and Stawarz, {\L}ukasz and Sugawara, Yasuharu and Sugita, Satoshi and Szymkowiak, Andrew and Tajima, Hiroyasu and Takahashi, Hiromitsu and Takahashi, Tadayuki and Takeda, Shin'ichiro and Takei, Yoh and Tamagawa, Toru and Tamura, Takayuki and Tanaka, Keigo and Tanaka, Takaaki and Tanaka, Yasuo and Tanaka, Yasuyuki T and Tashiro, Makoto S and Tawara, Yuzuru and Terada, Yukikatsu and Terashima, Yuichi and Tombesi, Francesco and Tomida, Hiroshi and Tsuboi, Yohko and Tsujimoto, Masahiro and Tsunemi, Hiroshi and Tsuru, Takeshi Go and Uchida, Hiroyuki and Uchiyama, Hideki and Uchiyama, Yasunobu and Ueda, Shutaro and Ueda, Yoshihiro and Uno, Shin'ichiro and Urry, C Megan and Ursino, Eugenio and Wang, Qian H S and Watanabe, Shin and Werner, Norbert and Wilkins, Dan R and Williams, Brian J and Yamada, Shinya and Yamaguchi, Hiroya and Yamaoka, Kazutaka and Yamasaki, Noriko Y and Yamauchi, Makoto and Yamauchi, Shigeo and Yaqoob, Tahir and Yatsu, Yoichi and Yonetoku, Daisuke and Zhuravleva, Irina and Zoghbi, Abderahmen},
  year = 2018,
  month = mar,
  journal = {Publications of the Astronomical Society of Japan},
  volume = {70},
  number = {2},
  pages = {9},
  issn = {0004-6264, 2053-051X},
  doi = {10.1093/pasj/psx138},
  urldate = {2025-05-09},
  langid = {english}
}

@article{Horton2020,
  title = {{{3D}} Hydrodynamic Simulations of Large-Scale Precessing Jets: Radio Morphology},
  shorttitle = {{{3D}} Hydrodynamic Simulations of Large-Scale Precessing Jets},
  author = {Horton, Maya A and Krause, Martin G H and Hardcastle, Martin J},
  year = 2020,
  month = nov,
  journal = {Monthly Notices of the Royal Astronomical Society},
  volume = {499},
  number = {4},
  pages = {5765--5781},
  issn = {0035-8711, 1365-2966},
  doi = {10.1093/mnras/staa3020},
  urldate = {2024-08-27},
  copyright = {https://academic.oup.com/journals/pages/open\_access/funder\_policies/chorus/standard\_publication\_model},
  langid = {english}
}

@article{Jerrim2024,
  title = {Faraday Rotation as a Probe of Radio Galaxy Environment in {{RMHD AGN}} Jet Simulations},
  author = {Jerrim, L A and Shabala, S S and {Yates-Jones}, P M and Krause, M G H and Turner, R J and Anderson, C S and Stewart, G S C and Power, C and Rodman, P E},
  year = 2024,
  month = may,
  journal = {Monthly Notices of the Royal Astronomical Society},
  volume = {531},
  number = {2},
  pages = {2532--2550},
  issn = {0035-8711, 1365-2966},
  doi = {10.1093/mnras/stae1317},
  urldate = {2024-08-27},
  copyright = {https://creativecommons.org/licenses/by/4.0/},
  langid = {english}
}

@article{Jones1979,
  title = {Hot Gas in Elliptical Galaxies and the Formation of Head-Tail Radio Sources},
  author = {Jones, T. W. and Owen, F. N.},
  year = 1979,
  month = dec,
  journal = {The Astrophysical Journal},
  volume = {234},
  pages = {818},
  issn = {0004-637X, 1538-4357},
  doi = {10.1086/157561},
  urldate = {2025-04-09},
  langid = {english}
}

@article{Jones2005,
  title = {An Efficient Numerical Scheme for Simulating Particle Acceleration in Evolving Cosmic-Ray Modified Shocks},
  author = {Jones, T. W. and Kang, Hyesung},
  year = 2005,
  month = sep,
  journal = {Astroparticle Physics},
  volume = {24},
  number = {1},
  pages = {75--91},
  issn = {0927-6505},
  doi = {10.1016/j.astropartphys.2005.05.006},
  urldate = {2025-01-28}
}

@article{Lacki2010,
  title = {The {{Physics}} of the {{Far-infrared-Radio Correlation}}. {{II}}. {{Synchrotron Emission}} as a {{Star Formation Tracer}} in {{High-redshift Galaxies}}},
  author = {Lacki, Brian C. and Thompson, Todd A.},
  year = 2010,
  month = jul,
  journal = {The Astrophysical Journal},
  volume = {717},
  pages = {196--208},
  publisher = {IOP},
  issn = {0004-637X},
  doi = {10.1088/0004-637X/717/1/196},
  urldate = {2025-08-12}
}

@article{Ledlow1996,
  title = {20 {{CM VLA}} Survey of Abell Clusters of Galaxies. {{VI}}. {{Radio}}/Optical Luminosity Functions},
  author = {Ledlow, Michael J. and Owen, Frazer N.},
  year = 1996,
  month = jul,
  journal = {The Astronomical Journal},
  volume = {112},
  pages = {9},
  issn = {00046256},
  doi = {10.1086/117985},
  urldate = {2025-02-25},
  langid = {english}
}

@article{Mathews2006,
  title = {Heating Cooling Flows with Weak Shock Waves},
  author = {Mathews, William G. and Faltenbacher, Andreas and Brighenti, Fabrizio},
  year = 2006,
  month = feb,
  journal = {The Astrophysical Journal},
  volume = {638},
  number = {2},
  pages = {659--667},
  issn = {0004-637X, 1538-4357},
  doi = {10.1086/499119},
  urldate = {2025-01-06},
  langid = {english}
}

@article{Matthews2019,
  title = {Ultrahigh Energy Cosmic Rays from Shocks in the Lobes of Powerful Radio Galaxies},
  author = {Matthews, J H and Bell, A R and Blundell, K M and Araudo, A T},
  year = 2019,
  month = feb,
  journal = {Monthly Notices of the Royal Astronomical Society},
  volume = {482},
  number = {4},
  pages = {4303--4321},
  issn = {0035-8711},
  doi = {10.1093/mnras/sty2936},
  urldate = {2025-03-05}
}

@article{Matthews2020,
  title = {Particle Acceleration in Astrophysical Jets},
  author = {Matthews, James H. and Bell, Anthony R. and Blundell, Katherine M.},
  year = 2020,
  month = sep,
  journal = {New Astronomy Reviews},
  volume = {89},
  pages = {101543},
  issn = {1387-6473},
  doi = {10.1016/j.newar.2020.101543},
  urldate = {2025-07-01}
}

@article{Meenakshi2024,
  title = {A Comparative Study of Radio Signatures from Winds and Jets: Modelling Synchrotron Emission and Polarization},
  shorttitle = {A Comparative Study of Radio Signatures from Winds and Jets},
  author = {Meenakshi, Moun and Mukherjee, Dipanjan and Bodo, Gianluigi and Rossi, Paola and Harrison, Chris M},
  year = 2024,
  month = sep,
  journal = {Monthly Notices of the Royal Astronomical Society},
  volume = {533},
  number = {2},
  pages = {2213--2231},
  issn = {0035-8711},
  doi = {10.1093/mnras/stae1890},
  urldate = {2025-06-19}
}

@article{Mendygral2012,
  title = {Mhd Simulations of Active Galactic Nucleus Jets in a Dynamic Galaxy Cluster Medium},
  author = {Mendygral, P. J. and Jones, T. W. and Dolag, K.},
  year = 2012,
  month = may,
  journal = {The Astrophysical Journal},
  volume = {750},
  number = {2},
  pages = {166},
  issn = {0004-637X, 1538-4357},
  doi = {10.1088/0004-637X/750/2/166},
  urldate = {2025-02-04},
  langid = {english}
}

@article{Mingo2019,
  title = {Revisiting the Fanaroff--Riley Dichotomy and Radio-Galaxy Morphology with the {{LOFAR}} Two-Metre Sky Survey ({{LoTSS}})},
  author = {Mingo, B and Croston, J H and Hardcastle, M J and Best, P N and Duncan, K J and Morganti, R and Rottgering, H J A and Sabater, J and Shimwell, T W and Williams, W L and Brienza, M and Gurkan, G and Mahatma, V H and Morabito, L K and Prandoni, I and Bondi, M and Ineson, J and Mooney, S},
  year = 2019,
  month = sep,
  journal = {Monthly Notices of the Royal Astronomical Society},
  volume = {488},
  number = {2},
  pages = {2701--2721},
  issn = {0035-8711, 1365-2966},
  doi = {10.1093/mnras/stz1901},
  urldate = {2023-10-24},
  langid = {english}
}

@article{Miyoshi2010,
  title = {The {{HLLD}} Approximate Riemann Solver for Magnetospheric Simulation},
  author = {Miyoshi, Takahiro and Terada, Naoki and Matsumoto, Yosuke and Fukazawa, Keiichiro and Umeda, Takayuki and Kusano, Kanya},
  year = 2010,
  month = sep,
  journal = {IEEE Transactions on Plasma Science},
  volume = {38},
  number = {9},
  pages = {2236--2242},
  issn = {1939-9375},
  doi = {10.1109/TPS.2010.2057451},
  urldate = {2025-03-06}
}

@article{Mukherjee2021,
  title = {Simulating the Dynamics and Synchrotron Emission from Relativistic Jets {{II}}. {{Evolution}} of Non-Thermal Electrons},
  author = {Mukherjee, Dipanjan and Bodo, Gianluigi and Rossi, Paola and Mignone, Andrea and Vaidya, Bhargav},
  year = 2021,
  month = jun,
  journal = {Monthly Notices of the Royal Astronomical Society},
  volume = {505},
  number = {2},
  eprint = {2105.02836},
  primaryclass = {astro-ph},
  pages = {2267--2284},
  issn = {0035-8711, 1365-2966},
  doi = {10.1093/mnras/stab1327},
  urldate = {2023-10-24},
  archiveprefix = {arXiv},
  langid = {english}
}

@article{Muller2021,
  title = {Two Striking Head--Tail Galaxies in the Galaxy Cluster {{IIZW108}}: Insights into Transition to Turbulence, Magnetic Fields, and Particle Re-Acceleration},
  shorttitle = {Two Striking Head--Tail Galaxies in the Galaxy Cluster {{IIZW108}}},
  author = {M{\"u}ller, Ancla and Pfrommer, Christoph and Ignesti, Alessandro and Moretti, Alessia and Louren{\c c}o, Ana and Paladino, Rosita and Jaff{\'e}, Yara and Gitti, Myriam and Venturi, Tiziana and Gullieuszik, Marco and Poggianti, Bianca and Vulcani, Benedetta and Biviano, Andrea and Adebahr, Bj{\"o}rn and Dettmar, Ralf-J{\"u}rgen},
  year = 2021,
  month = oct,
  journal = {Monthly Notices of the Royal Astronomical Society},
  volume = {508},
  number = {4},
  pages = {5326--5344},
  issn = {0035-8711, 1365-2966},
  doi = {10.1093/mnras/stab2928},
  urldate = {2025-07-04},
  copyright = {https://academic.oup.com/journals/pages/open\_access/funder\_policies/chorus/standard\_publication\_model},
  langid = {english}
}

@article{Murgia2011,
  title = {Dying Radio Galaxies in Clusters},
  author = {Murgia, M. and Parma, P. and Mack, K.-H. and De Ruiter, H. R. and Fanti, R. and Govoni, F. and Tarchi, A. and Giacintucci, S. and Markevitch, M.},
  year = 2011,
  month = feb,
  journal = {Astronomy \& Astrophysics},
  volume = {526},
  pages = {A148},
  issn = {0004-6361, 1432-0746},
  doi = {10.1051/0004-6361/201015302},
  urldate = {2025-04-15},
  langid = {english}
}

@article{Navarro1996,
  title = {The Structure of Cold Dark Matter Halos},
  author = {Navarro, Julio F. and Frenk, Carlos S. and White, Simon D. M.},
  year = 1996,
  month = may,
  journal = {The Astrophysical Journal},
  volume = {462},
  pages = {563},
  publisher = {IOP},
  issn = {0004-637X},
  doi = {10.1086/177173},
  urldate = {2025-01-06}
}

@article{Navarro1997,
  title = {A Universal Density Profile from Hierarchical Clustering},
  author = {Navarro, Julio F. and Frenk, Carlos S. and White, Simon D. M.},
  year = 1997,
  month = dec,
  journal = {The Astrophysical Journal},
  volume = {490},
  number = {2},
  pages = {493},
  publisher = {IOP Publishing},
  issn = {0004-637X},
  doi = {10.1086/304888},
  urldate = {2025-01-06},
  langid = {english}
}

@article{Nishikawa2020,
  title = {Rapid Particle Acceleration Due to Recollimation Shocks and Turbulent Magnetic Fields in Injected Jets with Helical Magnetic Fields},
  author = {Nishikawa, Kenichi and Mizuno, Yosuke and G{\'o}mez, Jose L and Du{\c t}an, Ioana and Niemiec, Jacek and Kobzar, Oleh and MacDonald, Nicholas and Meli, Athina and Pohl, Martin and Hirotani, Kouichi},
  year = 2020,
  month = apr,
  journal = {Monthly Notices of the Royal Astronomical Society},
  volume = {493},
  number = {2},
  pages = {2652--2658},
  issn = {0035-8711},
  doi = {10.1093/mnras/staa421},
  urldate = {2025-07-01}
}

@article{ONeill2019,
  title = {A Fresh Look at Narrow-Angle Tail Radio Galaxy Dynamics, Evolution, and Emissions},
  author = {O'Neill, Brian J. and Jones, T. W. and Nolting, Chris and Mendygral, P. J.},
  year = 2019,
  month = oct,
  journal = {The Astrophysical Journal},
  volume = {884},
  number = {1},
  pages = {12},
  issn = {1538-4357},
  doi = {10.3847/1538-4357/ab40b1},
  urldate = {2023-04-11},
  langid = {english}
}

@article{Pakmor2011,
  title = {Magnetohydrodynamics on an Unstructured Moving Grid: {{MHD}} on an Unstructured Moving Grid},
  shorttitle = {Magnetohydrodynamics on an Unstructured Moving Grid},
  author = {Pakmor, Ruediger and Bauer, Andreas and Springel, Volker},
  year = 2011,
  month = dec,
  journal = {Monthly Notices of the Royal Astronomical Society},
  volume = {418},
  number = {2},
  pages = {1392--1401},
  issn = {00358711},
  doi = {10.1111/j.1365-2966.2011.19591.x},
  urldate = {2025-01-10},
  langid = {english}
}

@article{Pakmor2013,
  title = {Simulations of Magnetic Fields in Isolated Disc Galaxies},
  author = {Pakmor, R{\"u}diger and Springel, Volker},
  year = 2013,
  month = jun,
  journal = {Monthly Notices of the Royal Astronomical Society},
  volume = {432},
  number = {1},
  pages = {176--193},
  issn = {1365-2966, 0035-8711},
  doi = {10.1093/mnras/stt428},
  urldate = {2025-01-17},
  langid = {english}
}

@ARTICLE{Whittingham2021,
       author = {{Whittingham}, Joseph and {Sparre}, Martin and {Pfrommer}, Christoph and {Pakmor}, R{\"u}diger},
        title = "{The impact of magnetic fields on cosmological galaxy mergers - I. Reshaping gas and stellar discs}",
      journal = {\mnras},
         year = 2021,
        month = sep,
       volume = {506},
       number = {1},
        pages = {229-255},
          doi = {10.1093/mnras/stab1425},
archivePrefix = {arXiv},
       eprint = {2011.13947},
 primaryClass = {astro-ph.GA},
       adsurl = {https://ui.adsabs.harvard.edu/abs/2021MNRAS.506..229W}
}

@article{Pakmor2016,
  title = {Semi-Implicit Anisotropic Cosmic Ray Transport on an Unstructured Moving Mesh},
  author = {Pakmor, R{\"u}diger and Pfrommer, Christoph and Simpson, Christine M. and Kannan, Rahul and Springel, Volker},
  year = 2016,
  month = nov,
  journal = {Monthly Notices of the Royal Astronomical Society},
  volume = {462},
  number = {3},
  pages = {2603--2616},
  issn = {0035-8711, 1365-2966},
  doi = {10.1093/mnras/stw1761},
  urldate = {2025-03-06},
  langid = {english}
}

@article{Perrone2022a,
  title = {Magneto-Thermal Instability in Galaxy Clusters -- {{I}}. {{Theory}} and Two-Dimensional Simulations},
  author = {Perrone, Lorenzo M and Latter, Henrik},
  year = 2022,
  month = may,
  journal = {Monthly Notices of the Royal Astronomical Society},
  volume = {513},
  number = {3},
  pages = {4605--4624},
  issn = {0035-8711, 1365-2966},
  doi = {10.1093/mnras/stac974},
  urldate = {2023-06-23},
  langid = {english}
}

@article{Peterson2006,
  title = {X-Ray Spectroscopy of Cooling Clusters},
  author = {Peterson, J. R. and Fabian, A. C.},
  year = 2006,
  month = apr,
  journal = {Physics Reports},
  volume = {427},
  number = {1},
  eprint = {astro-ph/0512549},
  pages = {1--39},
  issn = {03701573},
  doi = {10.1016/j.physrep.2005.12.007},
  urldate = {2023-04-11},
  archiveprefix = {arXiv},
  langid = {english}
}

@article{Pfrommer2010,
  title = {Detecting the Orientation of Magnetic Fields in Galaxy Clusters},
  author = {Pfrommer, Christoph and Dursi, L. J.},
  year = 2010,
  month = jul,
  journal = {Nature Physics},
  volume = {6},
  pages = {520--526},
  issn = {1745-2473},
  doi = {10.1038/nphys1657},
  urldate = {2025-07-01}
}

@article{Pfrommer2013,
  title = {Toward a Comprehensive Model for Feedback by Active Galactic Nuclei: New Insights from {{M87}} Observations by {{LOFAR}}, Fermi and h.e.s.s},
  shorttitle = {Toward a Comprehensive Model for Feedback by Active Galactic Nuclei},
  author = {Pfrommer, Christoph},
  year = 2013,
  month = nov,
  journal = {The Astrophysical Journal},
  volume = {779},
  number = {1},
  eprint = {1303.5443},
  primaryclass = {astro-ph},
  pages = {10},
  issn = {0004-637X, 1538-4357},
  doi = {10.1088/0004-637X/779/1/10},
  urldate = {2023-04-11},
  archiveprefix = {arXiv},
  langid = {english}
}

@article{Pfrommer2017,
  title = {Simulating Cosmic Ray Physics on a Moving Mesh},
  author = {Pfrommer, C. and Pakmor, R. and Schaal, K. and Simpson, C. M. and Springel, V.},
  year = 2017,
  month = mar,
  journal = {Monthly Notices of the Royal Astronomical Society},
  volume = {465},
  number = {4},
  eprint = {1604.07399},
  primaryclass = {astro-ph},
  pages = {4500--4529},
  issn = {0035-8711, 1365-2966},
  doi = {10.1093/mnras/stw2941},
  urldate = {2023-04-11},
  archiveprefix = {arXiv},
  langid = {english}
}

@article{Pfrommer2022,
  title = {Simulating Radio Synchrotron Emission in Star-Forming Galaxies: Small-Scale Magnetic Dynamo and the Origin of the Far-Infrared--Radio Correlation},
  shorttitle = {Simulating Radio Synchrotron Emission in Star-Forming Galaxies},
  author = {Pfrommer, Christoph and Werhahn, Maria and Pakmor, R{\"u}diger and Girichidis, Philipp and Simpson, Christine M},
  year = 2022,
  month = aug,
  journal = {Monthly Notices of the Royal Astronomical Society},
  volume = {515},
  number = {3},
  pages = {4229--4264},
  issn = {0035-8711, 1365-2966},
  doi = {10.1093/mnras/stac1808},
  urldate = {2025-07-01},
  copyright = {https://academic.oup.com/journals/pages/open\_access/funder\_policies/chorus/standard\_publication\_model},
  langid = {english}
}

@article{Pinzke2013,
  title = {Giant Radio Relics in Galaxy Clusters: Reacceleration of Fossil Relativistic Electrons?},
  shorttitle = {Giant Radio Relics in Galaxy Clusters},
  author = {Pinzke, Anders and Oh, S. Peng and Pfrommer, Christoph},
  year = 2013,
  month = oct,
  journal = {Monthly Notices of the Royal Astronomical Society},
  volume = {435},
  number = {2},
  eprint = {1301.5644},
  primaryclass = {astro-ph},
  pages = {1061--1082},
  issn = {1365-2966, 0035-8711},
  doi = {10.1093/mnras/stt1308},
  urldate = {2023-10-19},
  archiveprefix = {arXiv},
  langid = {english}
}

@article{Powell1999,
  title = {A Solution-Adaptive Upwind Scheme for Ideal Magnetohydrodynamics},
  author = {Powell, Kenneth G. and Roe, Philip L. and Linde, Timur J. and Gombosi, Tamas I. and De Zeeuw, Darren L.},
  year = 1999,
  month = sep,
  journal = {Journal of Computational Physics},
  volume = {154},
  number = {2},
  pages = {284--309},
  issn = {0021-9991},
  doi = {10.1006/jcph.1999.6299},
  urldate = {2025-03-06}
}

@article{Rafferty2006,
  title = {The Feedback-regulated Growth of Black Holes and Bulges through Gas Accretion and Starbursts in Cluster Central Dominant Galaxies},
  author = {Rafferty, D. A. and McNamara, B. R. and Nulsen, P. E. J. and Wise, M. W.},
  year = 2006,
  month = nov,
  journal = {The Astrophysical Journal},
  volume = {652},
  number = {1},
  pages = {216--231},
  issn = {0004-637X, 1538-4357},
  doi = {10.1086/507672},
  urldate = {2024-04-29},
  langid = {english}
}

@article{Ruszkowski2007,
  title = {Impact of Tangled Magnetic Fields on Fossil Radio Bubbles},
  author = {Ruszkowski, M. and Ensslin, T. A. and Bruggen, M. and Heinz, S. and Pfrommer, C.},
  year = 2007,
  month = jun,
  journal = {Monthly Notices of the Royal Astronomical Society},
  volume = {378},
  number = {2},
  pages = {662--672},
  issn = {0035-8711, 1365-2966},
  doi = {10.1111/j.1365-2966.2007.11801.x},
  urldate = {2025-03-05},
  langid = {english}
}

@article{Ruszkowski2017,
  title = {Cosmic-Ray Feedback Heating of the Intracluster Medium},
  author = {Ruszkowski, Mateusz and Yang, H.-Y. Karen and Reynolds, Christopher S.},
  year = 2017,
  month = jul,
  journal = {The Astrophysical Journal},
  volume = {844},
  number = {1},
  pages = {13},
  issn = {0004-637X, 1538-4357},
  doi = {10.3847/1538-4357/aa79f8},
  urldate = {2024-11-18},
  langid = {english}
}

@misc{Ruszkowski2023,
  title = {Cosmic Ray Feedback in Galaxies and Galaxy Clusters -- a Pedagogical Introduction and a Topical Review of the Acceleration, Transport, Observables, and Dynamical Impact of Cosmic Rays},
  author = {Ruszkowski, Mateusz and Pfrommer, Christoph},
  year = 2023,
  month = jun,
  number = {arXiv:2306.03141},
  eprint = {2306.03141},
  primaryclass = {astro-ph, physics:physics},
  publisher = {arXiv},
  urldate = {2023-06-14},
  archiveprefix = {arXiv}
}

@book{Rybicki1986,
  title = {Radiative Processes in Astrophysics},
  author = {Rybicki, George B. and Lightman, Alan P.},
  year = 1986,
  month = jun,
  journal = {Radiative Processes in Astrophysics},
  publisher = {Wiley},
  address = {New York, NY},
  urldate = {2025-04-28}
}

@article{Sabater2019,
  title = {The {{LoTSS}} View of Radio {{AGN}} in the Local Universe: {{The}} Most Massive Galaxies Are Always Switched On},
  shorttitle = {The {{LoTSS}} View of Radio {{AGN}} in the Local Universe},
  author = {Sabater, J. and Best, P. N. and Hardcastle, M. J. and Shimwell, T. W. and Tasse, C. and Williams, W. L. and Br{\"u}ggen, M. and Cochrane, R. K. and Croston, J. H. and De Gasperin, F. and Duncan, K. J. and G{\"u}rkan, G. and Mechev, A. P. and Morabito, L. K. and Prandoni, I. and R{\"o}ttgering, H. J. A. and Smith, D. J. B. and Harwood, J. J. and Mingo, B. and Mooney, S. and Saxena, A.},
  year = 2019,
  month = feb,
  journal = {Astronomy \& Astrophysics},
  volume = {622},
  pages = {A17},
  issn = {0004-6361, 1432-0746},
  doi = {10.1051/0004-6361/201833883},
  urldate = {2025-04-17},
  copyright = {https://www.edpsciences.org/en/authors/copyright-and-licensing},
  langid = {english}
}

@article{Sarazin1999,
  title = {The Energy Spectrum of Primary Cosmic-Ray Electrons in Clusters of Galaxies and Inverse Compton Emission},
  author = {Sarazin, Craig L.},
  year = 1999,
  month = aug,
  journal = {The Astrophysical Journal},
  volume = {520},
  number = {2},
  pages = {529},
  publisher = {IOP Publishing},
  issn = {0004-637X},
  doi = {10.1086/307501},
  urldate = {2023-08-15},
  langid = {english}
}

@article{Schoenmakers2000,
  title = {Radio Galaxies with a 'Double-Double Morphology'- {{I}}. {{Analysis}} of the Radio Properties and Evidence for Interrupted Activity in Active Galactic Nuclei},
  author = {Schoenmakers, A. P. and De Bruyn, A. G. and Rottgering, H. J. A. and Van Der Laan, H. and Kaiser, C. R.},
  year = 2000,
  month = jun,
  journal = {Monthly Notices of the Royal Astronomical Society},
  volume = {315},
  number = {2},
  pages = {371--380},
  issn = {0035-8711, 1365-2966},
  doi = {10.1046/j.1365-8711.2000.03430.x},
  urldate = {2025-02-25},
  langid = {english}
}

@article{Sijacki2008,
  title = {Simulations of Cosmic-Ray Feedback by Active Galactic Nuclei in Galaxy Clusters},
  author = {Sijacki, Debora and Pfrommer, Christoph and Springel, Volker and Enlin, Torsten A.},
  year = 2008,
  month = jul,
  journal = {Monthly Notices of the Royal Astronomical Society},
  volume = {387},
  number = {4},
  pages = {1403--1415},
  issn = {00358711, 13652966},
  doi = {10.1111/j.1365-2966.2008.13310.x},
  urldate = {2025-06-19},
  langid = {english}
}

@article{Sironi2014,
  title = {Relativistic Reconnection: {{An}} Efficient Source of Non-Thermal Particles},
  shorttitle = {Relativistic Reconnection},
  author = {Sironi, Lorenzo and Spitkovsky, Anatoly},
  year = 2014,
  month = feb,
  journal = {The Astrophysical Journal Letters},
  volume = {783},
  number = {1},
  pages = {L21},
  publisher = {The American Astronomical Society},
  issn = {2041-8205},
  doi = {10.1088/2041-8205/783/1/L21},
  urldate = {2025-07-01},
  langid = {english}
}

@article{Sironi2021,
  title = {Reconnection-Driven Particle Acceleration in Relativistic Shear Flows},
  author = {Sironi, Lorenzo and Rowan, Michael E. and Narayan, Ramesh},
  year = 2021,
  month = feb,
  journal = {The Astrophysical Journal Letters},
  volume = {907},
  number = {2},
  pages = {L44},
  issn = {2041-8205, 2041-8213},
  doi = {10.3847/2041-8213/abd9bc},
  urldate = {2025-04-15},
  langid = {english}
}

@article{Springel2010,
  title = {E Pur Si Muove: {{Galiliean-invariant}} Cosmological Hydrodynamical Simulations on a Moving Mesh},
  shorttitle = {E Pur Si Muove},
  author = {Springel, Volker},
  year = 2010,
  month = jan,
  journal = {Monthly Notices of the Royal Astronomical Society},
  volume = {401},
  number = {2},
  eprint = {0901.4107},
  primaryclass = {astro-ph},
  pages = {791--851},
  issn = {00358711, 13652966},
  doi = {10.1111/j.1365-2966.2009.15715.x},
  urldate = {2023-04-11},
  archiveprefix = {arXiv},
  langid = {english}
}

@article{Su2012,
  title = {{{EVIDENCE FOR GAMMA-RAY JETS IN THE MILKY WAY}}},
  author = {Su, Meng and Finkbeiner, Douglas P.},
  year = 2012,
  month = jul,
  journal = {The Astrophysical Journal},
  volume = {753},
  number = {1},
  pages = {61},
  issn = {0004-637X, 1538-4357},
  doi = {10.1088/0004-637X/753/1/61},
  urldate = {2025-07-04},
  langid = {english}
}

@article{Su2021,
  title = {Which {{AGN}} Jets Quench Star Formation in Massive Galaxies?},
  author = {Su, Kung-Yi and Hopkins, Philip F and Bryan, Greg L and Somerville, Rachel S and Hayward, Christopher C and {Angl{\'e}s-Alc{\'a}zar}, Daniel and {Faucher-Gigu{\`e}re}, Claude-Andr{\'e} and Wellons, Sarah and Stern, Jonathan and Terrazas, Bryan A and Chan, T K and Orr, Matthew E and Hummels, Cameron and Feldmann, Robert and Kere{\v s}, Du{\v s}an},
  year = 2021,
  month = aug,
  journal = {Monthly Notices of the Royal Astronomical Society},
  volume = {507},
  number = {1},
  pages = {175--204},
  issn = {0035-8711, 1365-2966},
  doi = {10.1093/mnras/stab2021},
  urldate = {2025-06-20},
  copyright = {https://academic.oup.com/journals/pages/open\_access/funder\_policies/chorus/standard\_publication\_model},
  langid = {english}
}

@article{Vaidya2018,
  title = {A Particle Module for the {{PLUTO}} Code. {{II}}. {{Hybrid}} Framework for Modeling Nonthermal Emission from Relativistic Magnetized Flows},
  author = {Vaidya, Bhargav and Mignone, Andrea and Bodo, Gianluigi and Rossi, Paola and Massaglia, Silvano},
  year = 2018,
  month = oct,
  journal = {The Astrophysical Journal},
  volume = {865},
  number = {2},
  pages = {144},
  issn = {1538-4357},
  doi = {10.3847/1538-4357/aadd17},
  urldate = {2023-10-24},
  langid = {english}
}

@article{vanWeeren2019,
  title = {Diffuse Radio Emission from Galaxy Clusters},
  author = {{van Weeren}, R. J. and {de Gasperin}, F. and Akamatsu, H. and Br{\"u}ggen, M. and Feretti, L. and Kang, H. and Stroe, A. and Zandanel, F.},
  year = 2019,
  month = feb,
  journal = {Space Science Reviews},
  volume = {215},
  number = {1},
  eprint = {1901.04496},
  primaryclass = {astro-ph},
  pages = {16},
  issn = {0038-6308, 1572-9672},
  doi = {10.1007/s11214-019-0584-z},
  urldate = {2023-04-11},
  archiveprefix = {arXiv},
  langid = {english}
}

@article{Vazza2021,
  title = {Simulating the Transport of Relativistic Electrons and Magnetic Fields Injected by Radio Galaxies in the Intracluster Medium},
  author = {Vazza, F. and Wittor, D. and Brunetti, G. and Br{\"u}ggen, M.},
  year = 2021,
  month = sep,
  journal = {Astronomy \& Astrophysics},
  volume = {653},
  eprint = {2102.04193},
  primaryclass = {astro-ph},
  pages = {A23},
  issn = {0004-6361, 1432-0746},
  doi = {10.1051/0004-6361/202140513},
  urldate = {2023-04-11},
  archiveprefix = {arXiv},
  langid = {english}
}

@article{Wang2020,
  title = {Chaotic Cold Accretion in Giant Elliptical Galaxies Heated by {{AGN}} Cosmic Rays},
  author = {Wang, Chaoran and Ruszkowski, Mateusz and Yang, H-Y Karen},
  year = 2020,
  month = apr,
  journal = {Monthly Notices of the Royal Astronomical Society},
  volume = {493},
  number = {3},
  pages = {4065--4076},
  issn = {0035-8711},
  doi = {10.1093/mnras/staa550},
  urldate = {2025-06-20}
}

@article{Wang2021,
  title = {Particle Acceleration in Shearing Flows: The Case for Large-Scale Jets},
  shorttitle = {Particle Acceleration in Shearing Flows},
  author = {Wang, Jie-Shuang and Reville, Brian and Liu, Ruo-Yu and Rieger, Frank M and Aharonian, Felix A},
  year = 2021,
  month = jun,
  journal = {Monthly Notices of the Royal Astronomical Society},
  volume = {505},
  number = {1},
  pages = {1334--1341},
  issn = {0035-8711, 1365-2966},
  doi = {10.1093/mnras/stab1458},
  urldate = {2025-07-01},
  copyright = {https://academic.oup.com/journals/pages/open\_access/funder\_policies/chorus/standard\_publication\_model},
  langid = {english}
}

@article{Weinberger2017,
  title = {Simulating the Interaction of Jets with the Intracluster Medium},
  author = {Weinberger, Rainer and Ehlert, Kristian and Pfrommer, Christoph and Pakmor, R{\"u}diger and Springel, Volker},
  year = 2017,
  month = oct,
  journal = {Monthly Notices of the Royal Astronomical Society},
  volume = {470},
  number = {4},
  pages = {4530--4546},
  issn = {0035-8711, 1365-2966},
  doi = {10.1093/mnras/stx1409},
  urldate = {2023-04-11},
  langid = {english}
}

@article{Weinberger2020,
  title = {The Arepo Public Code Release},
  author = {Weinberger, Rainer and Springel, Volker and Pakmor, R{\"u}diger},
  year = 2020,
  month = jun,
  journal = {The Astrophysical Journal Supplement Series},
  volume = {248},
  number = {2},
  eprint = {1909.04667},
  primaryclass = {astro-ph, physics:physics},
  pages = {32},
  issn = {1538-4365},
  doi = {10.3847/1538-4365/ab908c},
  urldate = {2023-04-11},
  archiveprefix = {arXiv},
  langid = {english}
}

@article{Weinberger2023,
  title = {Active Galactic Nucleus Jet Feedback in Hydrostatic Haloes},
  author = {Weinberger, Rainer and Su, Kung-Yi and Ehlert, Kristian and Pfrommer, Christoph and Hernquist, Lars and Bryan, Greg L and Springel, Volker and Li, Yuan and Burkhart, Blakesley and Choi, Ena and {Faucher-Gigu{\`e}re}, Claude-Andr{\'e}},
  year = 2023,
  month = may,
  journal = {Monthly Notices of the Royal Astronomical Society},
  volume = {523},
  number = {1},
  pages = {1104--1125},
  issn = {0035-8711, 1365-2966},
  doi = {10.1093/mnras/stad1396},
  urldate = {2025-01-06},
  copyright = {https://academic.oup.com/journals/pages/open\_access/funder\_policies/chorus/standard\_publication\_model},
  langid = {english}
}

@article{Werhahn2021c,
  title = {Cosmic Rays and Non-Thermal Emission in Simulated Galaxies: {{III}}. Probing Cosmic Ray Calorimetry with Radio Spectra and the {{FIR-radio}} Correlation},
  shorttitle = {Cosmic Rays and Non-Thermal Emission in Simulated Galaxies},
  author = {Werhahn, Maria and Pfrommer, Christoph and Girichidis, Philipp},
  year = 2021,
  month = oct,
  journal = {Monthly Notices of the Royal Astronomical Society},
  volume = {508},
  number = {3},
  eprint = {2105.12134},
  primaryclass = {astro-ph},
  pages = {4072--4095},
  issn = {0035-8711, 1365-2966},
  doi = {10.1093/mnras/stab2535},
  urldate = {2023-04-11},
  archiveprefix = {arXiv},
  langid = {english}
}

@misc{Werhahn2025,
  title = {Steady-{{State}} or {{Not}}? {{The Evolution}} of {{Cosmic Ray Electron Spectra}} in {{Galaxies}}},
  shorttitle = {Steady-{{State}} or {{Not}}?},
  author = {Werhahn, Maria and Pfrommer, Christoph and Whittingham, Joseph and Jlassi, L{\'e}na and Pakmor, R{\"u}diger and Girichidis, Philipp and Bieri, Rebekka},
  year = 2025,
  month = nov,
  number = {arXiv:2511.13811},
  eprint = {2511.13811},
  primaryclass = {astro-ph},
  publisher = {arXiv},
  doi = {10.48550/arXiv.2511.13811},
  urldate = {2025-12-05},
  archiveprefix = {arXiv},
  langid = {english}
}

@misc{Whittingham2024,
  title = {Zooming-in on Cluster Radio Relics -- {{I}}. {{How}} Density Fluctuations Explain the Mach Number Discrepancy, Microgauss Magnetic Fields, and Spectral Index Variations},
  author = {Whittingham, Joseph and Pfrommer, Christoph and Werhahn, Maria and Jlassi, L{\'e}na and Girichidis, Philipp},
  year = 2024,
  month = nov,
  number = {arXiv:2411.11947},
  eprint = {2411.11947},
  primaryclass = {astro-ph},
  publisher = {arXiv},
  doi = {10.48550/arXiv.2411.11947},
  urldate = {2025-01-10},
  archiveprefix = {arXiv},
  langid = {english}
}

@article{Winner2019,
  title = {Evolution of Cosmic Ray Electron Spectra in Magnetohydrodynamical Simulations},
  author = {Winner, Georg and Pfrommer, Christoph and Girichidis, Philipp and Pakmor, R{\"u}diger},
  year = 2019,
  month = sep,
  journal = {Monthly Notices of the Royal Astronomical Society},
  volume = {488},
  number = {2},
  eprint = {1903.01467},
  primaryclass = {astro-ph},
  pages = {2235--2252},
  issn = {0035-8711, 1365-2966},
  doi = {10.1093/mnras/stz1792},
  urldate = {2023-04-11},
  archiveprefix = {arXiv},
  langid = {english}
}

@article{Winner2020,
  title = {Evolution and Observational Signatures of the Cosmic Ray Electron Spectrum in {{SN}} 1006},
  author = {Winner, Georg and Pfrommer, Christoph and Girichidis, Philipp and Werhahn, Maria and Pais, Matteo},
  year = 2020,
  month = oct,
  journal = {Monthly Notices of the Royal Astronomical Society},
  volume = {499},
  number = {2},
  eprint = {2006.06683},
  primaryclass = {astro-ph},
  pages = {2785--2802},
  issn = {0035-8711, 1365-2966},
  doi = {10.1093/mnras/staa2989},
  urldate = {2023-04-11},
  archiveprefix = {arXiv},
  langid = {english}
}

@article{Yang2017,
  title = {The Spatially Uniform Spectrum of the Fermi Bubbles: {{The}} Leptonic Active Galactic Nucleus Jet Scenario},
  shorttitle = {The Spatially Uniform Spectrum of the Fermi Bubbles},
  author = {Yang, H.-Y. K. and Ruszkowski, M.},
  year = 2017,
  month = nov,
  journal = {The Astrophysical Journal},
  volume = {850},
  number = {1},
  pages = {2},
  publisher = {The American Astronomical Society},
  issn = {0004-637X},
  doi = {10.3847/1538-4357/aa9434},
  urldate = {2025-06-19},
  langid = {english}
}

@article{Yang2019a,
  title = {Extended Catalog of Winged or X-Shaped Radio Sources from the {{FIRST}} Survey},
  author = {Yang, Xiaolong and Joshi, Ravi and {Gopal-Krishna} and An, Tao and Ho, Luis C. and Wiita, Paul J. and Liu, Xiang and Yang, Jun and Wang, Ran and Wu, Xue-Bing and Yang, Xiaofeng},
  year = 2019,
  month = nov,
  journal = {The Astrophysical Journal Supplement Series},
  volume = {245},
  number = {1},
  pages = {17},
  issn = {0067-0049, 1538-4365},
  doi = {10.3847/1538-4365/ab4811},
  urldate = {2025-04-10},
  langid = {english}
}

@article{Yates-Jones2022,
  title = {{{PRAiSE}}: Resolved Spectral Evolution in Simulated Radio Sources},
  shorttitle = {{{PRAiSE}}},
  author = {{Yates-Jones}, Patrick M and Turner, Ross J and Shabala, Stanislav S and Krause, Martin G H},
  year = 2022,
  month = apr,
  journal = {Monthly Notices of the Royal Astronomical Society},
  volume = {511},
  number = {4},
  pages = {5225--5240},
  issn = {0035-8711},
  doi = {10.1093/mnras/stac385},
  urldate = {2025-06-19}
}





\begin{appendix}

\section{Cooling timescales}\label{sec:appendix_cooling_times}

\begin{figure}[ht]
	\centering
	\includegraphics[width=1.\columnwidth]{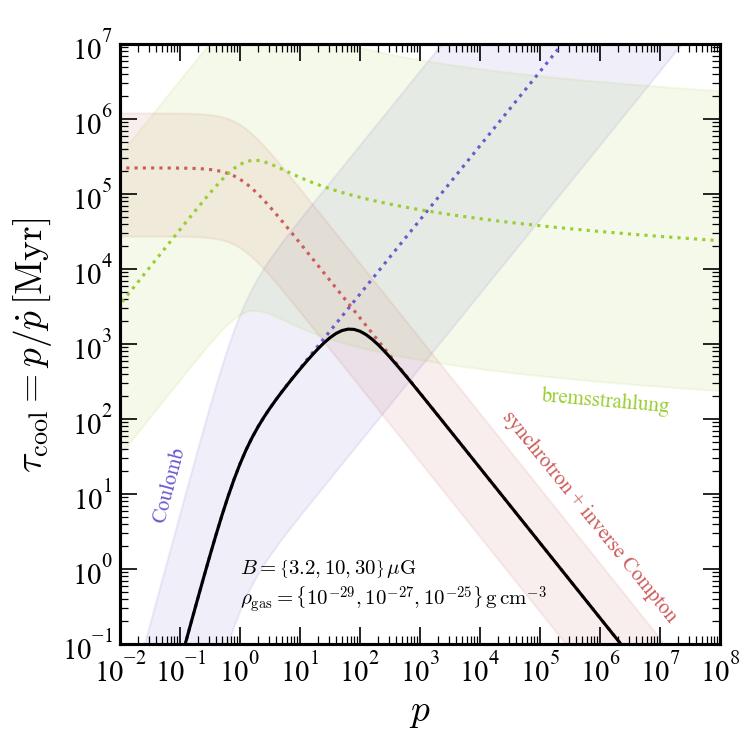}

	\caption{Cooling timescales as a function of electron momentum for Coulomb (blue), combined synchrotron and inverse Compton (red), and bremsstrahlung (green) losses. We show these loss timescales for a range of gas densities and magnetic field strengths. The minimum and maximum values of these quantities correspond to the edge curves of each distribution, while the dotted coloured line is the intermediate value. The black solid curve combines the cooling timescales for all processes using the intermediate value. The value of $3.2\,\mu \rm{G}$ corresponds to the CMB equivalent magnetic field. At low momenta, cooling timescales can be shorter than 200~Myr in denser environments such as in the ICM. At high momenta, cooling timescales shorter than 200~Myr occur in regions of high magnetic field strengths such as in the jet or in the wake of jet-inflated lobes.
	}
    \label{fig:app_cooling_times}
\end{figure}

In Fig.~\ref{fig:app_cooling_times}, we show a range of cooling time curves against the normalized momentum $p$. We show Coulomb collisions which dominate at low momenta, bremsstrahlung losses, and synchrotron and inverse Compton losses which dominate at high momenta. We select a range of gas densities between $10^{-29} \leq \rho_{\rm{gas}} / \rm{g} \,\rm{cm}^{-3} \leq 10^{-25}$, and a range of magnetic field values between $3.2 \leq B / \mu \rm{G} \leq 30$, relevant for the environments in which CRe populations evolve in our simulations. We choose the minimum value of $B = 3.2\, \mu \rm{G}$ corresponding to the CMB-equivalent magnetic field, which is an upper limit to the cooling timescales at high momenta: CRes will be cooled by the CMB-equivalent magnetic field through inverse Compton scattering even when they encounter lower magnetic field strengths. The selected mass density is appropriately converted to electron and gas number densities according to the parameters listed in Sect.~\ref{subsubsec:cre}.

\section{Scrutinizing adiabatic effects on proton and electron distributions}\label{sec:appendix_CRep_distributions}

\begin{figure*}[]
	\centering
	\includegraphics[width=1.3\columnwidth]{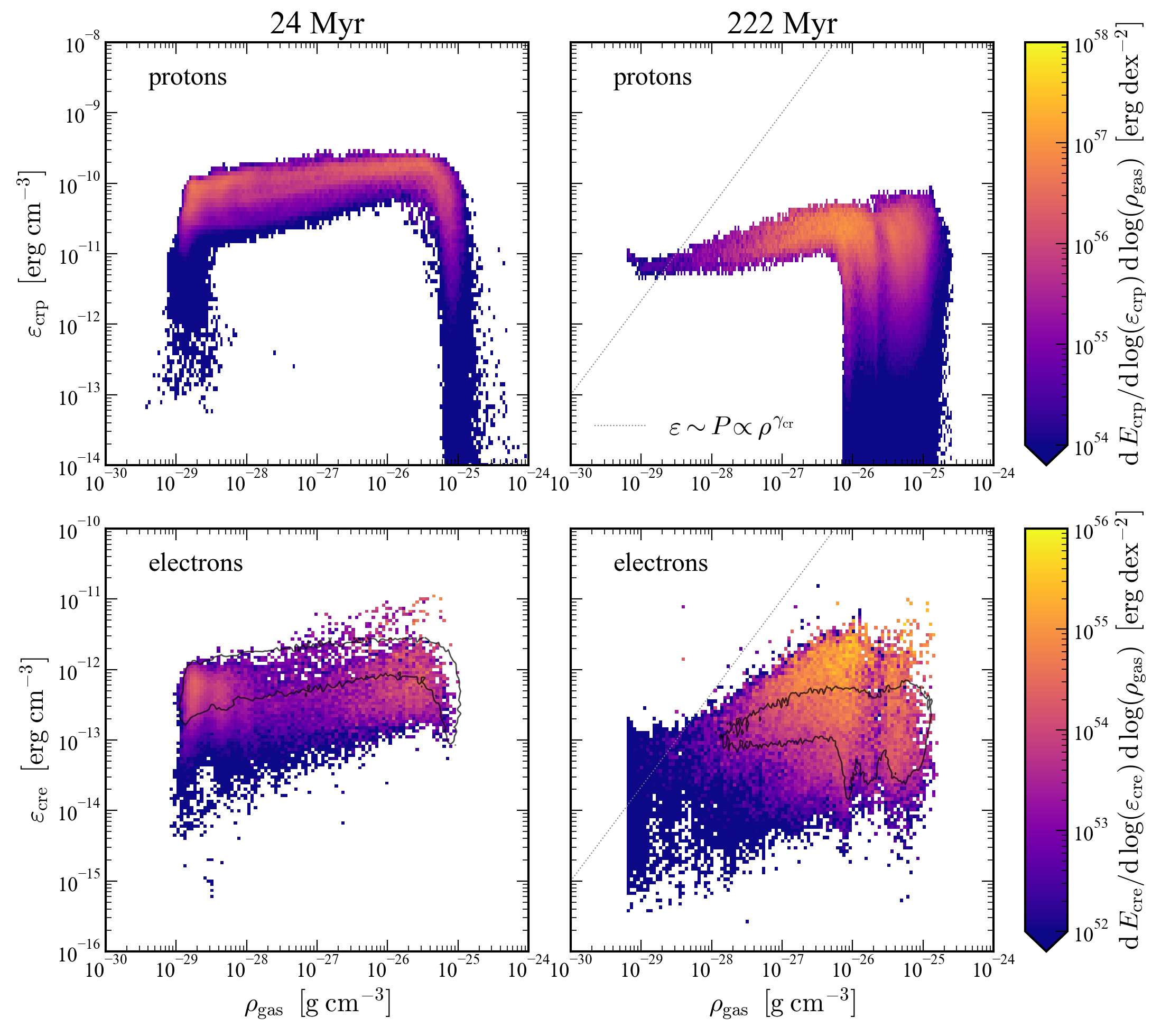}
	\caption{Two-dimensional histograms of CR energy density against the gas density, using models that only include adiabatic changes and mixing. Each pixel is colour-coded by the total CR energy per bin. We show protons (top) and electrons (bottom) halfway through jet activity (left panels) and at the end of the simulation (right panels). The axis limits for the CRe energy density and total energy are scaled by the CRe acceleration efficiency $\xi_{\rm{cre}} = 0.01$, for easier comparison between the two populations. Black contours on the bottom panel show $1\sigma$ of the CRp energy distribution, also scaled by the CRe acceleration efficiency. The relation between energy density and density upon adiabatic changes is shown as a dotted grey line. Most differences between the two populations can be explained by differences in discretization.
	\textit{Top:} CRp energy distributions which display tails at low and high densities due to mixing of jet material with the ICM. CRps are distributed across a range of densities due to mixing and to the progressive nature of our acceleration algorithm. The bulk of the energy moves to higher densities and deviates from the adiabat due to mixing.
	\textit{Bottom:} CRe energy distributions display a larger spread in energy density, but are distributed across a range of densities for the same reasons as the CRps. The bulk of the energy moves to higher densities, and follows the adiabat more closely.}
	\label{fig:app_crep_phase_diagrams}
	\centering
	\includegraphics[width=1.7\columnwidth]{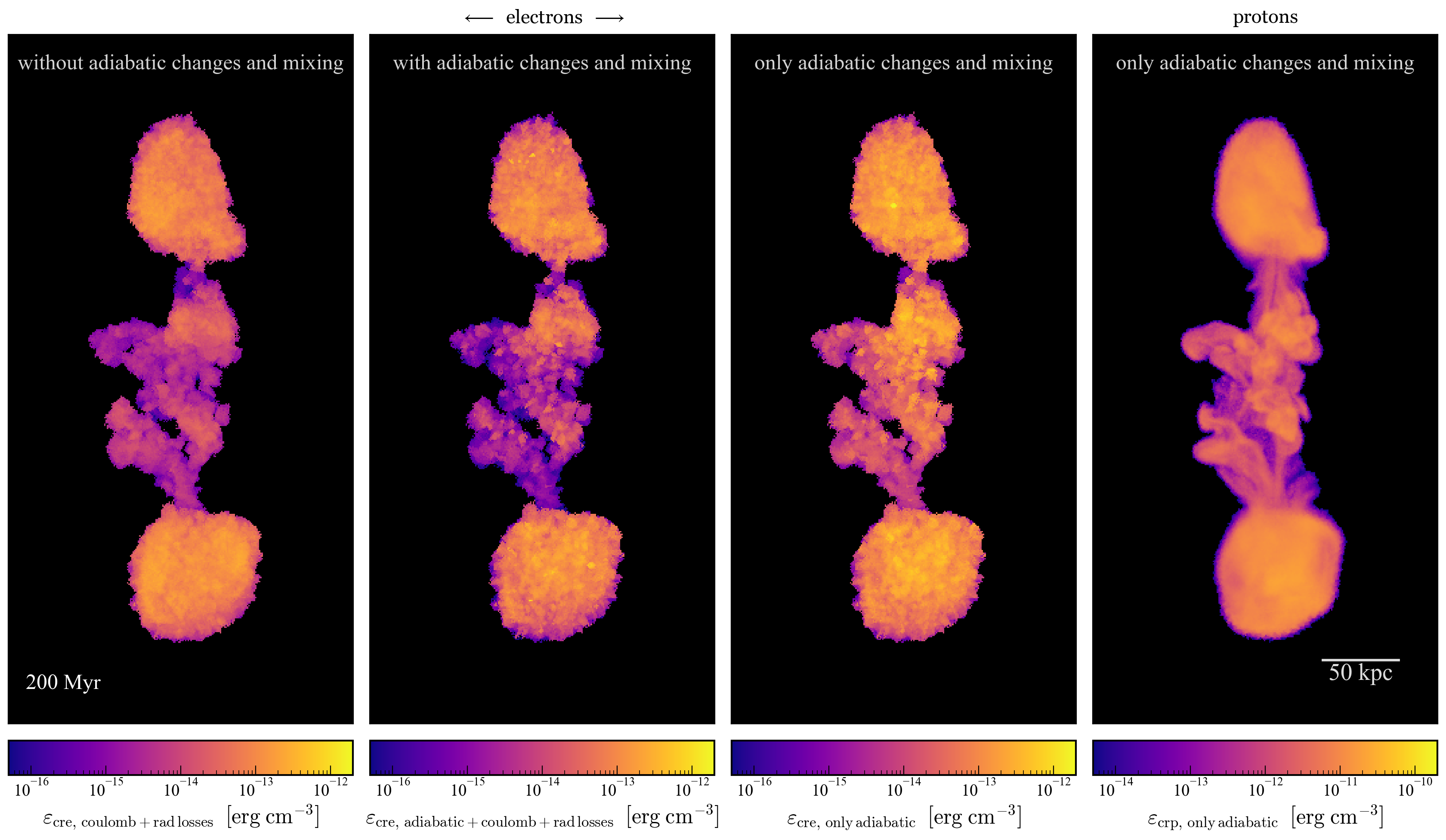}
	\caption{Projections of the volume-weighted CRe and CRp energy density with different models, from left to right: CRe without adiabatic changes and mixing, CRe experience Coulomb and radiative cooling as well as adiabatic changes and mixing, CRes with only adiabatic changes and mixing, purely advective CRp with only adiabatic changes (and mixing). All projections have a depth $\pm$ 60~kpc from the cluster centre.}
	\label{fig:app_crep_projections}
\end{figure*}

In Fig.~\ref{fig:app_crep_phase_diagrams}, we plot 2D distributions in energy density $\varepsilon_{\rm{cr}}$ and gas mass density $\rho_{\rm{gas}}$ weighted by the CR energy, for both CRps and CRes. We show these distributions at $t=24$~Myr, while the jet is active, and at the end of the simulation, at $t=222$~Myr, to show the effect of adiabatic effects on these populations. In the case of CRps, they are advected in the \textsc{Arepo} simulation, taking into account mass fluxes through cell interfaces and adiabatic effects. For the CRes, the adiabatic changes are performed using velocity field tracers, which do not follow mass flows perfectly \citep{Genel2013}. The model for adiabatic changes along these tracer particles is described in Sect.~\ref{subsec:adiabatic_effects}, where the gas density is used to track compression/expansion events, and the jet tracer $X_{\rm{jet}}$ is used to track the dilution of CRes through mixing of the jet material with the ICM.

Initially, the distribution of CRps (top-left panel in Fig.~\ref{fig:app_crep_phase_diagrams}) shows that most of the energy lies between densities of $10^{-29}$ and $10^{-25}~\mathrm{g~cm}^{-3}$. This widening of the distribution to densities larger than the jet density points to two effects occurring simultaneously. A Lagrangian element starting at the jet base and rising in the cluster atmosphere causes entrainment of dense ICM gas, leading to jet cells with a range of densities. This effect, combined with the progressive CRp acceleration algorithm, implies that CRp are injected in jet cells with a range of densities. The electron distribution (bottom left) roughly follows the CRp distribution but is more bimodal, with energy preferentially located at low and high densities. This is likely a consequence of the different discretization methods. Specifically, while cells with CRp energy mix with the external medium through mass fluxes, tracer particles with CRe energy do not, and instead partially capture this mixing by recording the mixed state of the cell they are located in (their parent cell). This effect is stronger in diverging flows, which occur in the jet `hotspots' and the wake of the lobes in our simulations. We defer the reader to \citet{Genel2013}, where the velocity tracers are discussed in greater detail. At late times, both distributions show that (CRp, CRe) energy has moved to higher densities. Neither distribution evolves along the adiabat -- they rather move along shallower slopes, hinting at mixing processes causing CRs to move along a downward diagonal. This is stronger in the case of CRps, which resolve mixing directly, while CRes resolve mixing `passively'. We also notice that while the CRp energy density has decreased overall (moving from $10^{-10}$ to $10^{-11}$), it has not decreased as much in the electron case. This is reflected in Fig.~\ref{fig:app_crep_projections}, which shows the spatial distribution of energy density in the jet at late times.

In Fig.~\ref{fig:app_crep_projections}, we show projections of the electron and proton energy density at 200~Myr. The rightmost panel is a projection of the \textsc{Arepo} Voronoi mesh, while the three leftmost panels are produced using the Voronoi mesh generated from tracer particle positions in post-processing. Comparing the two rightmost panels, we see that our treatment for adiabatic changes of CRes allows us to reproduce the general behaviour of CRps. There are local differences between the two components, which are due to the different discretizations. Specifically, the CRe energy density appears slightly larger than the CRp energy density in the wake of the bubbles due to converging flows towards the cluster centre. Turning to the second panel from the left, where Coulomb and radiative losses are included on top of adiabatic and mixing terms, the most significant difference is in the wake of the bubbles, where the energy density is lower by almost an order of magnitude compared to the panel with only adiabatic changes and mixing. Both the density and the magnetic field strength are higher in these regions, as shown in Fig.~\ref{fig:gas_projections}, hinting at either Coulomb or synchrotron loss processes, or the combination of both, being responsible for the loss of CRe energy in these central parts. Finally, the leftmost panel shows a model where only Coulomb and radiative losses are included. The energy density appears smoother in this panel in comparison to models where adiabatic changes and mixing are included (second and third panel from the left), highlighting the effect of adiabatic changes and mixing on local CRe populations. This is in agreement with the total electron spectrum with adiabatic changes shown in Fig.~\ref{fig:cre_spectrum}, which exhibits features of compression at momenta $p \sim 10$ and $p \sim 10^6$.

\end{appendix}

\end{document}